\documentclass[onecolumn,floatfix,superscriptaddress,showpacs,showkeys,nofootinbib,preprint]{revtex4}
\usepackage{epsfig}
\usepackage{amssymb,latexsym,amsmath}
\newcommand{\eq}[1]{\begin{align} #1 \end{align}}
\begin{document}

\title{  Baryon Number and Electric Charge Fluctuations\\
in Pb+Pb Collisions at SPS energies }

\author{V.P. Konchakovski}
\affiliation{Bogolyubov Institute for Theoretical Physics, Kiev,
Ukraine}
\affiliation{Shevchenko National University, Kiev, Ukraine}
\affiliation{Frankfurt Institute for Advanced Studies, Frankfurt,
Germany}
\author{ M.I. Gorenstein}
\affiliation{Bogolyubov Institute
for Theoretical Physics, Kiev, Ukraine}
\affiliation{Frankfurt
Institute for Advanced Studies, Frankfurt, Germany}
\author{E.L. Bratkovskaya}
\affiliation{Frankfurt Institute for Advanced Studies, Frankfurt, Germany}
\author{H. St\"ocker}
\affiliation{Frankfurt Institute for Advanced Studies, Frankfurt,
Germany}
\affiliation{Institut f\"ur Theoretische Physik, Johann
Wolfgang Goethe Universit\"at,
 Frankfurt, Germany}

\begin{abstract}
Event-by-event fluctuations of the net baryon number and electric
charge in nucleus-nucleus collisions are studied in Pb+Pb at SPS
energies within the HSD transport model. We reveal an important role of
the fluctuations in the number of target nucleon participants. They
strongly influence all measured fluctuations even in the samples of
events with rather rigid centrality trigger. This fact can be used to
check different scenarios of nucleus-nucleus collisions by measuring
the multiplicity fluctuations as a function of collision centrality
in fixed kinematical regions of the projectile and target hemispheres.
The HSD results for the event-by-event fluctuations of electric charge
in central Pb+Pb collisions at 20, 30, 40, 80 and 158 A GeV are in a
good agreement with the NA49 experimental data and considerably larger
than expected in a quark-gluon plasma. This demonstrate that the
distortions of the initial fluctuations by the hadronization phase and, in
particular, by the final resonance decays dominate the observable
fluctuations.

\end{abstract}
\maketitle

\section{Introduction}

The aim of the present paper is to study the fluctuations of the net
baryon number and electric charge in nucleus-nucleus
(A+A) collisions at SPS energies. We use the HSD \cite{HSD} transport
approach which reproduces both the different particle multiplicities
and longitudinal differential rapidity distributions for central
collisions of Au+Au (or Pb+Pb)  from AGS to SPS energies rather well
\cite{Weber}.  
(see, e.g.,
Refs.~\cite{fluc1,fluc2,fluc3,fluc4,fluc5,fluc6,fluc7,fluc7a,fluc7b,fluc8}
and references therein) reveal a new physical information. The
fluctuations in A+A collisions are studied on an event-by-event
basis: a given observable is measured in each event and the
fluctuations are evaluated for a specially selected set of these
events.  We recall that the statistical model has been successfully used to
describe the data on hadron multiplicities in relativistic A+A
collisions (see, e.g., Ref.~\cite{stat1} and a recent review
\cite{BMST}) as well as in elementary particle collisions
\cite{stat2}. This gives rise to the question whether the
fluctuations, in particular the multiplicity fluctuations, do also
follow the statistical hadron-resonance gas results. Recently the
particle number fluctuations have been studied in different
statistical ensembles \cite{stat3}; the statistical fluctuations can
be closely related to phase transitions in QCD matter, with specific
signatures for 1-st and 2-nd order phase transitions as well as for
the critical point \cite{fluc4,fluc5}.

In addition to the statistical fluctuations the complicated
time evolution of A+A collisions generates {\it dynamical} fluctuations.
The fluctuations in the initial energy deposited inelastically in
the statistical system yield {\it dynamical} fluctuations  of all
macroscopic parameters, like the total entropy or strangeness
content. The observable consequences of the initial energy density
fluctuations are sensitive to the equation of state,
and can therefore be useful as signals for phase transitions
\cite{fluc8}.  Even when the data are obtained with a centrality
trigger the number of nucleons participating in inelastic
collisions still fluctuates considerably. In the language of
statistical mechanics, these fluctuations in the participant nucleon
number correspond to volume fluctuations. Secondary particle
multiplicities scale linearly with the volume, hence, volume
fluctuations translate directly to particle number fluctuations.

The present work is a continuation of our recent study  \cite{KGB}
where we have analyzed the charged particle number fluctuations in
Pb+Pb collisions at 158~AGeV within the UrQMD and HSD transport
approaches.  The net baryon number and electric charge event-by-event
fluctuations are studied in different rapidity regions of the
projectile and target hemispheres.

The paper is organized as follows. Section II presents the HSD results
for the fluctuations of the number of nucleon participants while
Sections III and IV give the net baryon number fluctuations and
electric charge fluctuations, respectively. In Section V we discuss the
fluctuations in the samples of most central collisions, Section VI
shows a comparison of our calculations with experimental data from the
NA49 Collaboration, whereas Section VII finally concludes the present
study.

\section{Fluctuations of the Number of Participants}

 In each A+A collision only a fraction of all 2$A$ nucleons interact.
These are called participant nucleons and are denoted as
$N_P^{proj}$ and $N_P^{targ}$ for the projectile and target nuclei,
respectively. The nucleons, which do not interact, are called the
projectile and target spectators, $N_S^{proj} = A - N_P^{proj}$ and
$N_S^{targ} = A - N_P^{targ}$.
The fluctuations in high energy A+A collisions are dominated by a
geometrical  variation of the impact parameter. However,
even for the fixed impact parameter the number of participants,
$N_P\equiv N_P^{proj}+N_P^{targ}$, fluctuates from event to event.
This is due to the fluctuations of the initial states of the
colliding nuclei and the probabilistic character of the interaction
process. The fluctuations of $N_P$ form usually a large and
uninteresting background. In order to minimize its contribution the NA49
Collaboration has selected samples of collisions with a fixed numbers of
projectile participants. This selection is possible due to a
measurement of $N_S^{proj}$ in each individual collision by a
calorimeter which covers the projectile fragmentation domain.
However, even in the samples with $N_P^{proj} = const$ the number of
target participants fluctuates considerably. Hence, an asymmetry
between projectile and target participants is introduced, i.e.
$N_P^{proj}$ is constant by constraint, whereas $N_P^{targ}$ fluctuates
independently.

\begin{figure}[ht!]
\epsfig{file=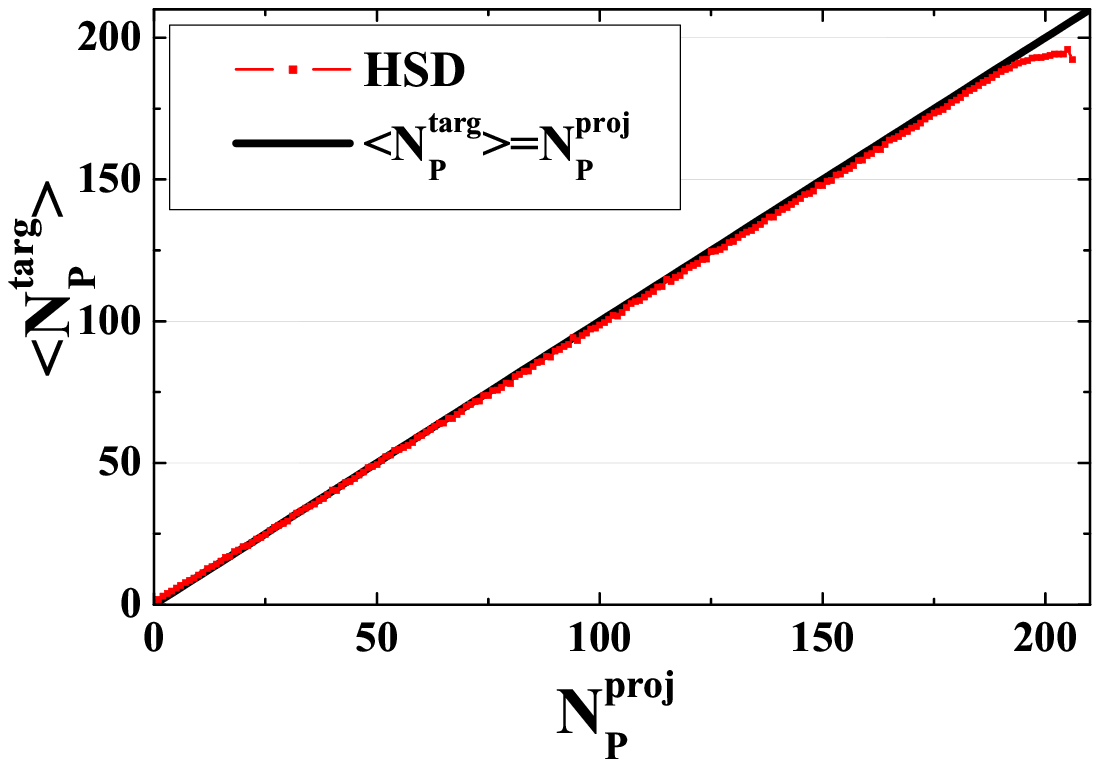,width=8cm}
 \epsfig{file=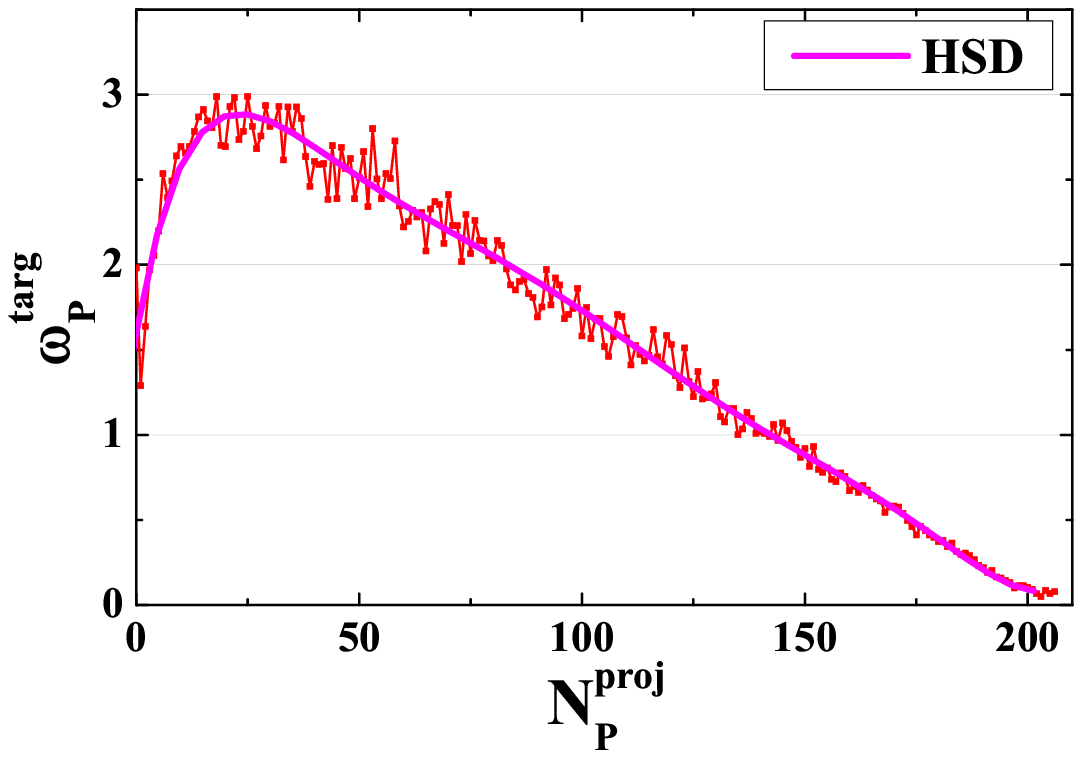,width=8cm}
\caption{ The HSD simulations in Pb+Pb collisions at 158~AGeV for the
average value $\langle N_P^{targ}\rangle$ ({\it left}) and the
scaled variances $\omega_P^{targ}$ ({\it right}) as
functions of $N_P^{proj}$. \label{wNtarg}}
\end{figure}

In the following the variance, $Var(n) \equiv \langle n^2 \rangle -
\langle n \rangle^2$, and scaled variance, $\omega_n \equiv
Var(n)/\langle n \rangle$, where $n$ stands for a given random
variable and $\langle \cdots \rangle$ for event-by-event averaging,
will be used to quantify fluctuations. In each sample with
$N_P^{proj}=const$ the number of target participants fluctuates
around its mean value, $\langle N_P^{targ} \rangle$ with the scaled
variance $\omega_P^{targ}$. From an output of the HSD minimum bias
simulations of Pb+Pb collisions at 158~AGeV we form the samples of
events with fixed values of $N_P^{proj}$. Fig.~\ref{wNtarg} presents
the HSD average value $\langle N_P^{targ}\rangle$ (left) and the
scaled variances $\omega_P^{targ}$ (right) as functions of
$N_P^{proj}$. One finds $\langle N_P^{targ} \rangle \simeq
N_P^{proj}$; the deviations are only seen at very small
($N_P^{proj}\approx 1$) and very large ($N_P^{proj}\approx A$)
numbers of projectile participants. The fluctuations of $N_
P^{targ}$ are quite strong: $\omega_P^{targ}
> 2$  at $N_{P}^{proj}=10 - 80$.

\begin{figure}[ht!]
\epsfig{file=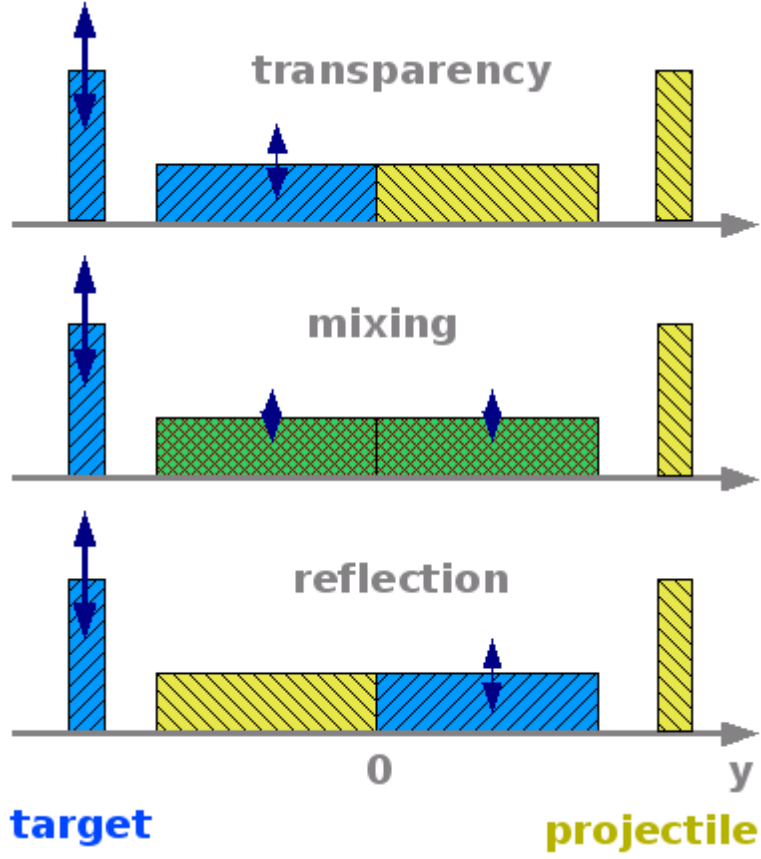,width=10cm} \caption{The sketch of the
rapidity distributions of the baryon number or the particle
production sources (horizontal rectangles) in nucleus-nucleus
collisions resulting from the transparency, mixing and reflection
models. The spectator nucleons are indicated by the vertical
rectangles. In the collisions with a fixed number of projectile
spectators only matter related to the target shows significant
fluctuations (vertical arrows). See Ref.~\cite{MGMG} for more
details. } \label{sketch}
\end{figure}

The consequences of the asymmetry between projectile and target
hemispheres depend on the A+A dynamics. According to
Ref.~\cite{MGMG} different models of hadron production in
relativistic A+A collisions can be divided into three limiting
groups: transparency (T-), mixing (M-), and reflection (R-)
models. The rapidity distributions resulting from the T-, M-, and
R-models are sketched in Fig.~\ref{sketch} taken from
Ref.~\cite{MGMG}. We note that there are models which assume
the mixing of hadron production sources, however,
the transparency of baryon flows, e.g. three-fluid
hydrodynamical model \cite{3fluid}.
R-models appear rather unrealistic and are included for completeness
in our discussion.

\section{Net Baryon Number Fluctuations}

We begin with a  quantitative discussion by first considering the
fluctuations of the net baryon number in different regions
of the participant domain in collisions of two identical nuclei.
These fluctuations are most closely related to the fluctuations of
the number of participant nucleons because of baryon number
conservation.

\begin{figure}[ht!]
\epsfig{file=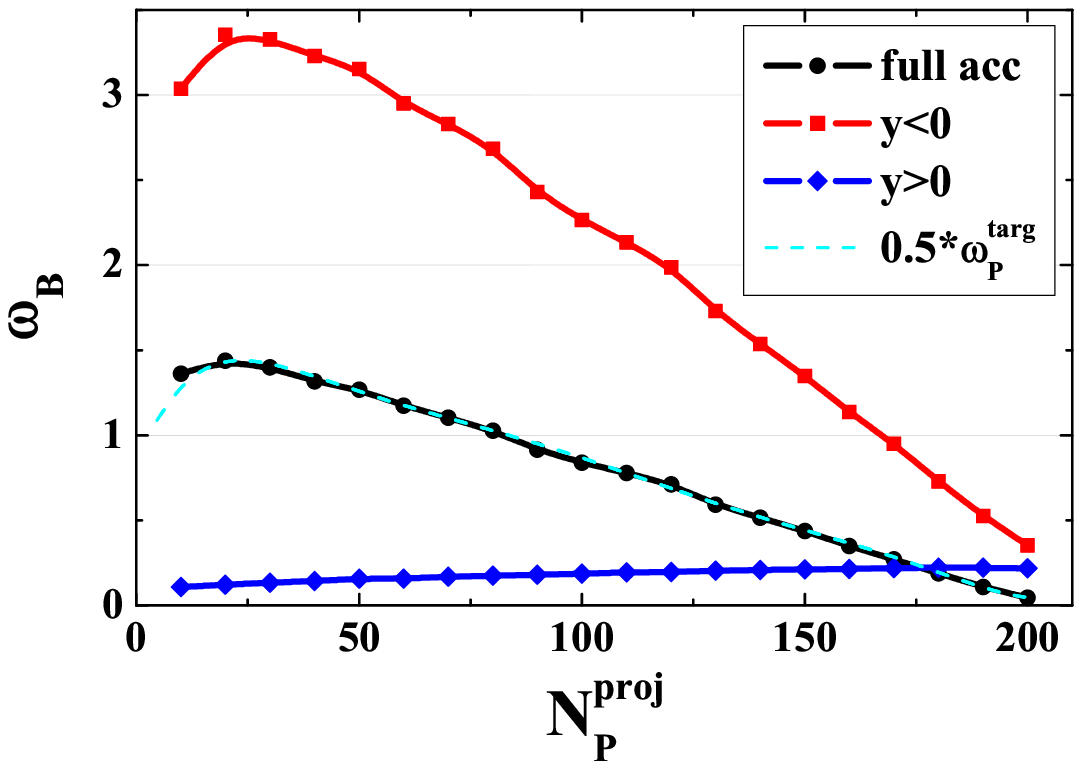,width=8cm}
\epsfig{file=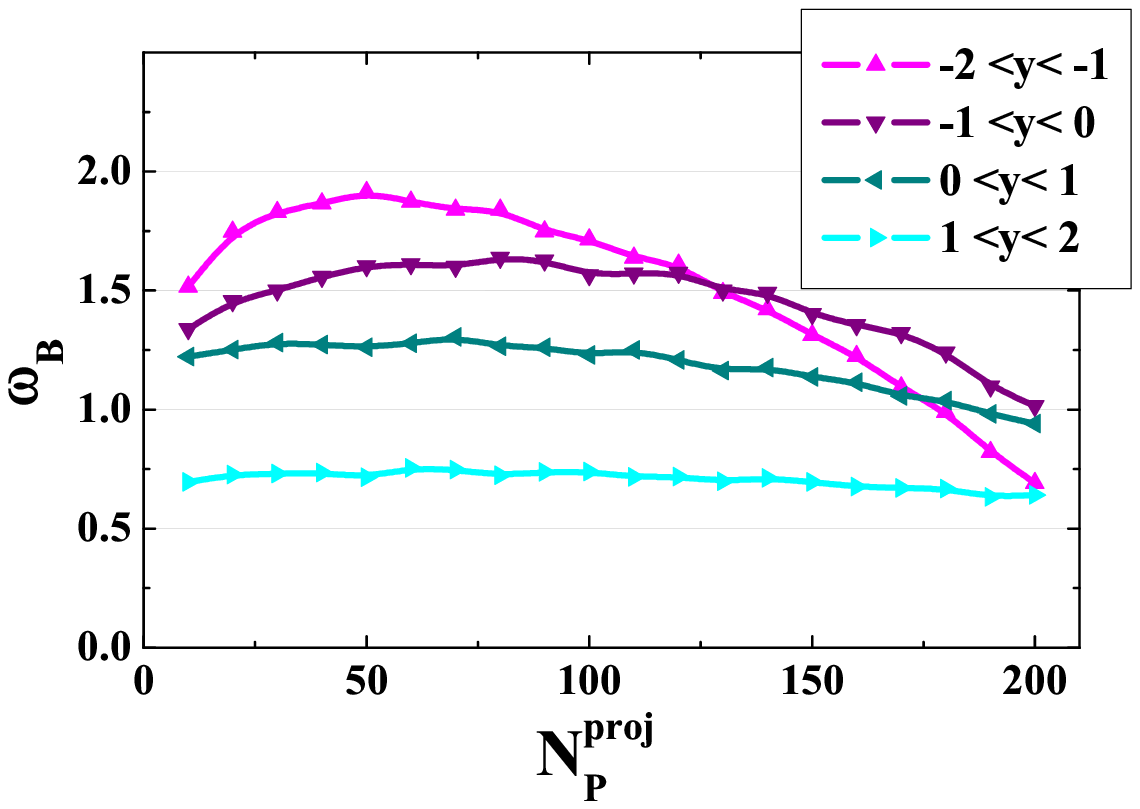,width=8cm}
\caption{ The HSD simulations for Pb+Pb collisions at 158~AGeV for
fixed values of $N_P^{proj}$. {\it Left:} The baryon number
fluctuations in full acceptance, $\omega_B$, in projectile
hemisphere, $\omega_B^p$ (lower curve), and in target hemisphere,
$\omega_B^t$ (upper curve). The dashed line, $0.5~\omega_P^{targ}$,
demonstrates the validity of the relation (\ref{omegaB}). {\it Right:}
The scaled variances of the baryon number fluctuations in different
rapidity intervals. \label{wB1}}
\end{figure}

The HSD results for $\omega_B$ in Pb+Pb at 158~AGeV are presented in
Fig.~\ref{wB1}.
In each event we subtract the nucleon spectators when counting the
number of baryons. The net baryon number in the full phase space,
$B\equiv N_B-N_{\overline{B}}$, equals then to the total number of
participants $N_P=N_P^{targ}+N_P^{proj}$. At fixed $N_P^{proj}$ the
$N_P$ number fluctuates due to  fluctuations of $N_P^{targ}$.
These fluctuations correspond to an average value, $\langle
N_P^{targ}\rangle \simeq N_P^{proj}$, and a scaled variance,
$\omega_P^{targ}$ (see Fig.~\ref{wNtarg}). Thus, for the net baryon
number fluctuations in the full phase space we find,
\eq{ \omega_B~=~\frac{Var(N_P)}{\langle N_P\rangle}~\simeq~
\frac{\langle \left(N_P^{targ}\right)^2\rangle~-~\langle
N_P^{targ}\rangle^2}{2\langle
N_P^{targ}\rangle}~=~\frac{1}{2}~\omega_P^{targ}~. \label{omegaB} }
A factor $1/2$ in the right hand side of Eq.~(\ref{omegaB}) appears
because only half of the total number of participants fluctuates.

Let us introduce $\omega_B^p$ and $\omega_B^t$, where the
superscripts $p$ and $t$ mark quantities measured in the
projectile and target momentum hemispheres, respectively.
Fig.~\ref{wB1} demonstrates that $\omega_B^t>\omega_B^p$, both in
the whole projectile-target hemispheres and in the symmetric
rapidity intervals. On the other hand one observes that
$\omega_B^p \approx \omega_B^t$ in most central collisions. This
is because the fluctuations of the target participants become
negligible in this case, i.e. $\omega_P^{targ}\rightarrow 0$
(Fig.~\ref{wNtarg}, right). As a consequence the fluctuations of
any observable in the symmetric rapidity intervals become
identical in most central collisions. Note also that
transparency-mixing effects are  different at different
rapidities. From Fig.~\ref{wNtarg} (right) it follows that
$\omega_B^p$ in the target rapidity interval $[-2,-1]$ is much
larger than $\omega_B^t$ in the symmetric projectile rapidity
interval $[1,2]$. This fact reveals the strong transparency
effects. On the other hand, the behavior is different in symmetric
rapidity intervals near the midrapidity. From
Fig.~\ref{wNtarg} (right) one observes that $\omega_B^p$ in the
target rapidity interval $[-1,0]$ is already much closer to
$\omega_B^t$ in the symmetric projectile rapidity interval
$[0,1]$. This gives a rough estimate of the width, $\Delta y
\approx 1$, for the region in rapidity space where projectile and
target nucleons communicate to each others.

By assumption, the mixing of the projectile and target participants
is absent in T- and R-models. Therefore,
in T-models, the net baryon number in the projectile hemisphere
equals to $N_p^{proj}$ and does not fluctuate, i.e.
$\omega_B^{p}(T)=0$, whereas the net baryon number in the target
hemisphere equals to $N_p^{targ}$ and fluctuates with
$\omega_B^{t}(T)=\omega_P^{targ}$.
These relations are reversed in R-models.
We introduce now a mixing of baryons between the projectile and
target hemispheres. Let $\alpha$ be the probability for a (projectile)
target participant to be detected in the (target) projectile
hemisphere. We denote by $n^t$ and $n^p$ the number of
baryons which end uo in the target and projectile hemisphere,
respectively, from the opposite hemisphere. Then the probabilities
to detect $B^t$ baryons in the target hemisphere, and $B^p$
baryons  in the projectile hemisphere, can be written as,
\eq{
& P(B^t;N_P^{proj})~=~\sum_{N_P^{targ}}W(N_P^{targ};
N_P^{proj})~\sum_{n^t=1}^{N_P^{targ}} \sum_{n^p=1}^{N_P^{proj}}
 \alpha^{n^p}(1-\alpha)^{N_P^{targ}-n^p}~
 \frac{N_P^{targ}!}{n^p!(N_P^{targ}-n^p)!} \nonumber \\
 & \times~
\alpha^{n^t}(1-\alpha)^{N_P^{proj}-n^t}~\frac{N_P^{proj}!}
{n^t!(N_P^{proj}-n^t)!}~\delta
\left(B^t-N_P^{targ}~-n^t~+~n^p\right)~, \label{PBt}\\
& P(B^p;N_P^{proj})~=~\sum_{N_P^{targ}}W(N_P^{targ};
N_P^{proj})~\sum_{n^t=1}^{N_P^{targ}} \sum_{n^p=1}^{N_P^{proj}}
 \alpha^{n^p}(1-\alpha)^{N_P^{targ}-n^p}~
 \frac{N_P^{targ}!}{n^p!(N_P^{targ}-n^p)!} \nonumber \\
 & \times~
\alpha^{n^t}(1-\alpha)^{N_P^{proj}-n^t}~\frac{N_P^{proj}!}
{n^t!(N_P^{proj}-n^t)!}~\delta
\left(B^p-N_P^{proj}~-~n^p~+~n^t\right)~,\label{PBp}
}
where $W(N_P^{targ}; N_P^{proj})$ is the probability distribution of
$N_P^{targ}$ in a sample with fixed value of $N_P^{proj}$. From
Eqs.~(\ref{PBt},\ref{PBp}) with a straightforward calculation we
find:
\eq{
\omega_B^t~=~(1~-~\alpha)^2~\omega_P^{targ}~+~2\alpha(1~-~\alpha)~,~~~~
\omega_B^p~=~\alpha^2~\omega_P^{targ}~+~2\alpha(1~-~\alpha)~.
\label{alpha1}
}
 A (complete) mixing of the projectile and target participants
is assumed in M-models. Thus each participant nucleon with equal
probability, $\alpha=1/2$, can be found either in the target or in
projectile hemispheres. In M-models the fluctuations in both
projectile and target hemispheres are identical. The limiting cases,
$\alpha=0$ and $\alpha=1$, of Eq.~(\ref{alpha1}) correspond to T-
and R-models, respectively. In summary, the scaled variances of the
net baryon number fluctuations in the projectile, $\omega_B^p$, and
target, $\omega_B^t$, hemispheres are:
\eq{ &\omega_B^{p}(T)~=~0~,~~~~~~
\omega_B^{t}(T)~=~\omega_P^{targ}~, \label{TB}\\
&\omega_B^{p}(M)~=~\omega_B^{t}(M)~=~\frac{1}{2}~+~
~\frac{1}{4}~ \omega_P^{targ}~,\label{MB}\\
&\omega_B^{p}(R)~=~\omega_P^{targ}~, ~~~~ \omega_B^{t}(R)~=~0~,
\label{RB} }
in the T- (\ref{TB}), M- (\ref{MB}) and R- (\ref{RB}) models of the
baryon number flow.
The different models lead to significantly different predictions for
$\omega_B^p$ and $\omega_B^t$.

\begin{figure}[ht!]
\epsfig{file=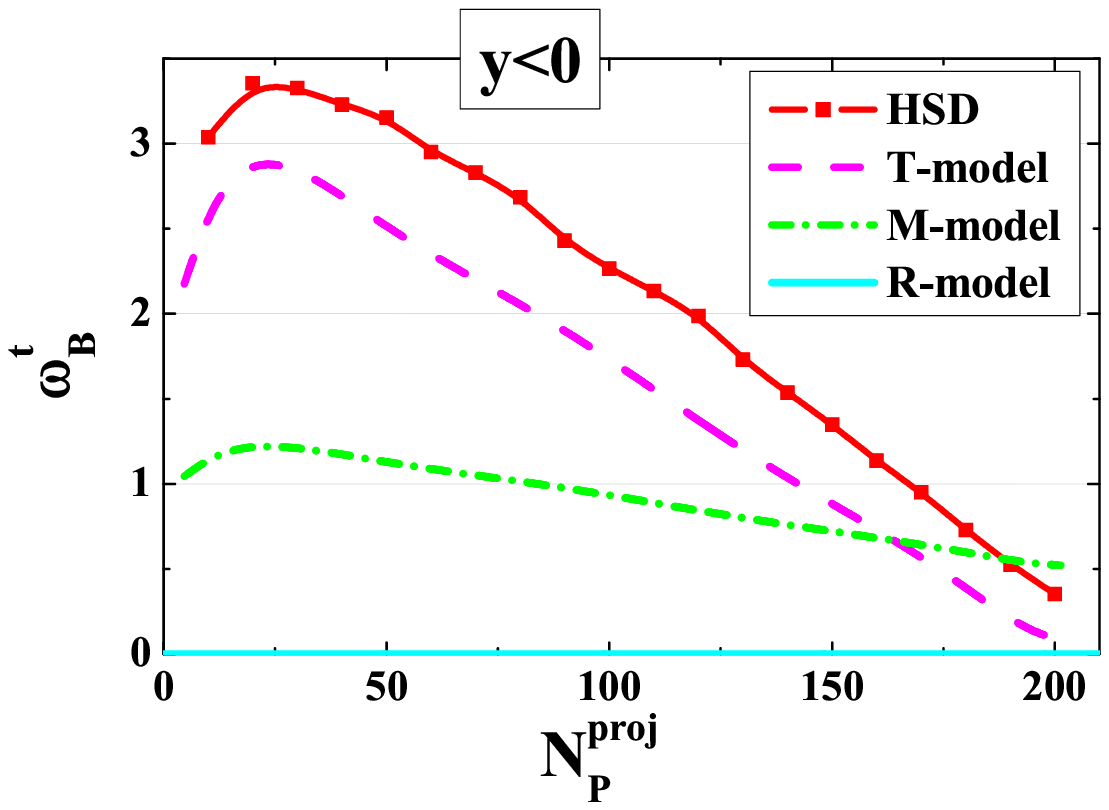,width=8cm}
\epsfig{file=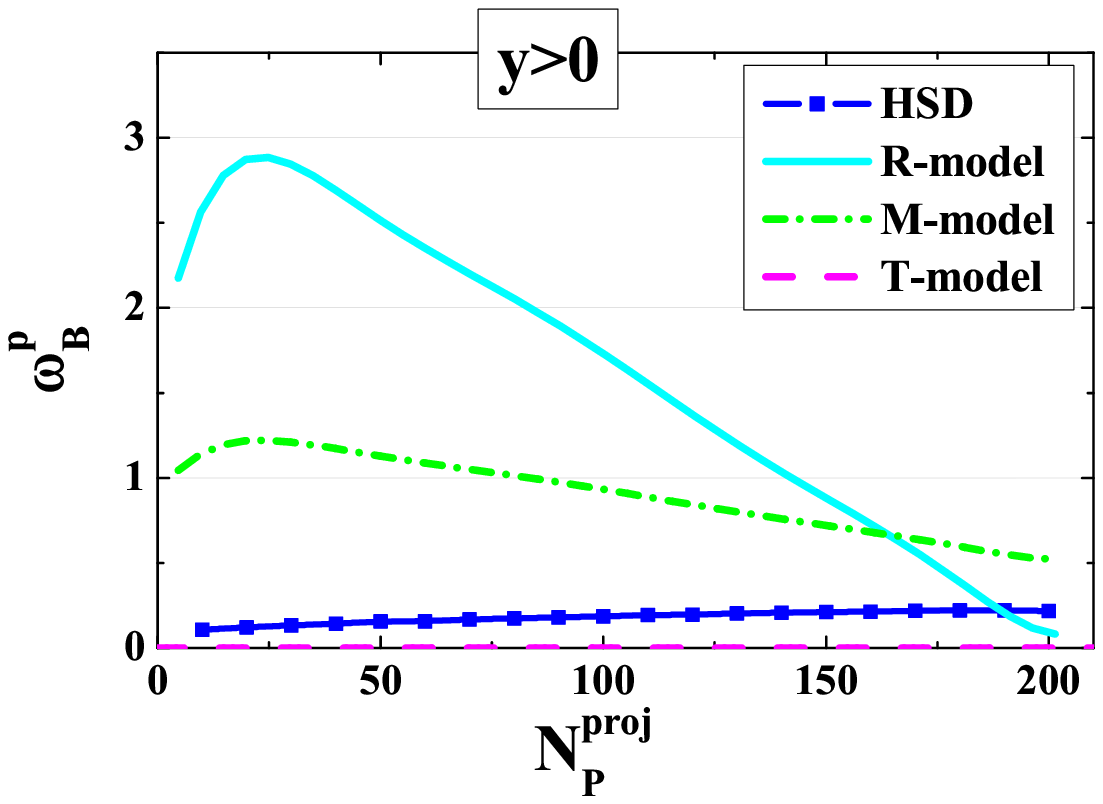,width=8cm} \caption{The $\omega_B^t$ ({\it
left}) and $\omega_B^p$ ({\it right}) of the HSD simulations in
comparison to T-, M- and R-models (\ref{TB}-\ref{RB}), with
$\omega_P^{targ}$ taken from Fig.~\ref{wNtarg}.} \label{wB_models}
\end{figure}

In Fig.~\ref{wB_models} we show the
predictions of T-, M- and R-models (\ref{TB}-\ref{RB}) with
$\omega_P^{targ}$ from Fig.~\ref{wNtarg}
for Pb+Pb collisions at 158$A$ GeV.
%
%
From Fig.~\ref{wB_models} one concludes that the HSD results are
close to the T-model estimates for baryon flow. However, the
deviations from the results (\ref{TB}) are clearly seen:
$\omega_B^p>0$ and $\omega_B^t>\omega_P^{targ}$. One can not fit the
HSD values of $\omega_B^t$ and $\omega_B^p$ by Eq.~(\ref{alpha1}).
To make $\omega_B^p>0$ one needs $\alpha >0$, but this induces
$\omega_B^t <\omega_P^{targ}$, i.e. a mixing of baryons between the
projectile and target hemispheres creates a non-zero baryon number
fluctuations in the projectile hemisphere on the expense of
fluctuations in the target hemisphere. Indeed, it follows from
Eq.~(\ref{alpha1}) that $\omega_B^p$ increases with $\alpha$ for all
$\alpha$, if $\omega_P^{targ}>1$, and for $\alpha <
(2-\omega_P^{targ})^{-1}$, if $\omega_P^{targ}<1$. On the other
hand, $\omega_B^t$ increases with $\alpha$ if $\alpha
<(1-\omega_P^{targ})(2 -\omega_P^{targ})^{-1}$. This shows that an
increase of $\omega_B^t$ with $\alpha$ is only possible for
$\omega_P^{targ}<1$. Thus for $\omega_P^{targ}>1$ one finds an
increase of $\omega_B^p$ with $\alpha$ and a decrease of
$\omega_B^t$ with $\alpha$ for all physical values of $\alpha$ from
0 to 1. Therefore, we conclude that the HSD values of
$\omega_B^t$ (i.e. the fact that $\omega_B^t>\omega_P^{targ}$) can
not be explained by Eq.~(\ref{alpha1}) with $\alpha >0$.

\begin{figure}[ht!]
\hspace*{-2cm} \epsfig{file=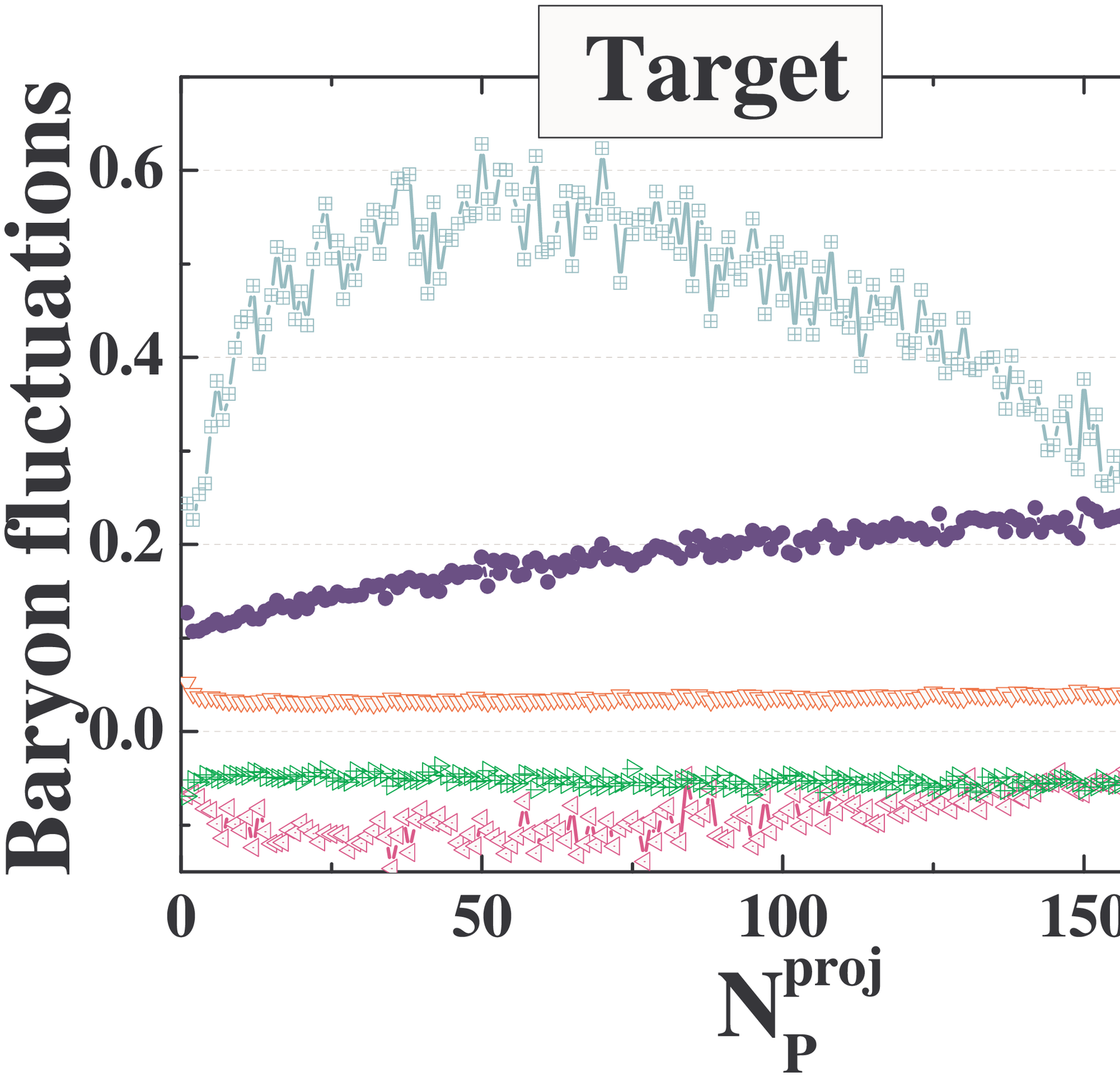,scale=0.25}
\hspace*{3cm} \epsfig{file=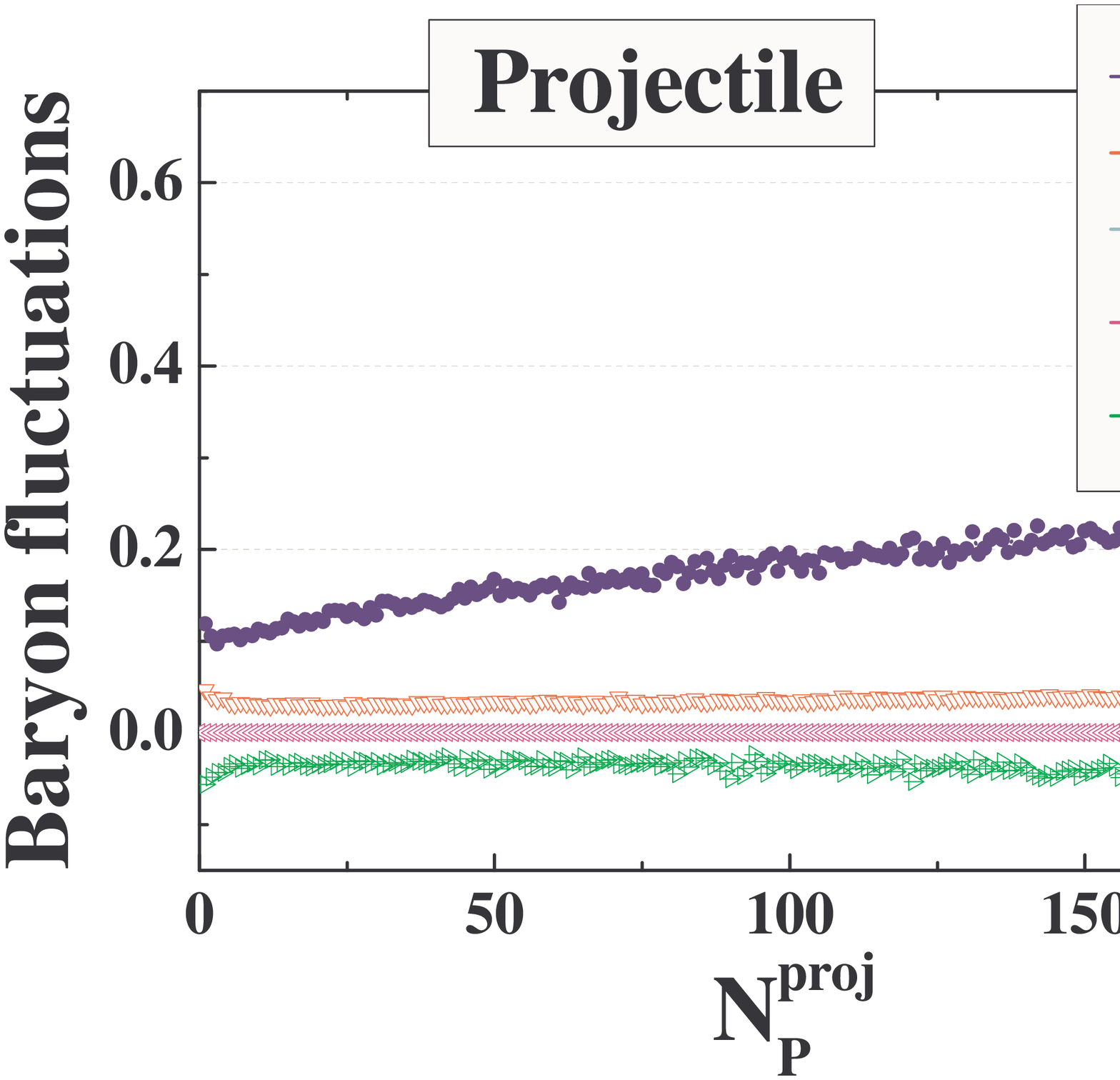,scale=0.25} \hfill
\caption{Different terms of
Eq.~(\ref{termsBt}), {\it left}, and Eq.~(\ref{termsBp}), {\it
right}, are presented as a function of $N_P^{proj}$. } \label{terms}
\end{figure}

\begin{figure}[ht!]
\epsfig{file=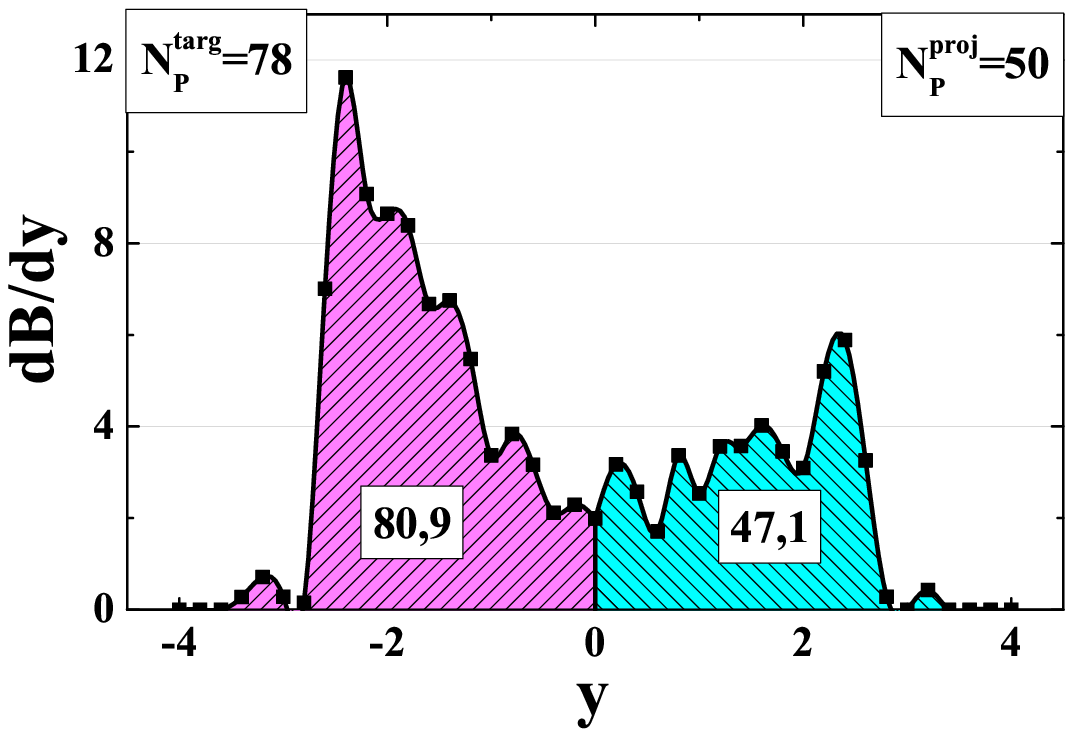,width=8cm}
\epsfig{file=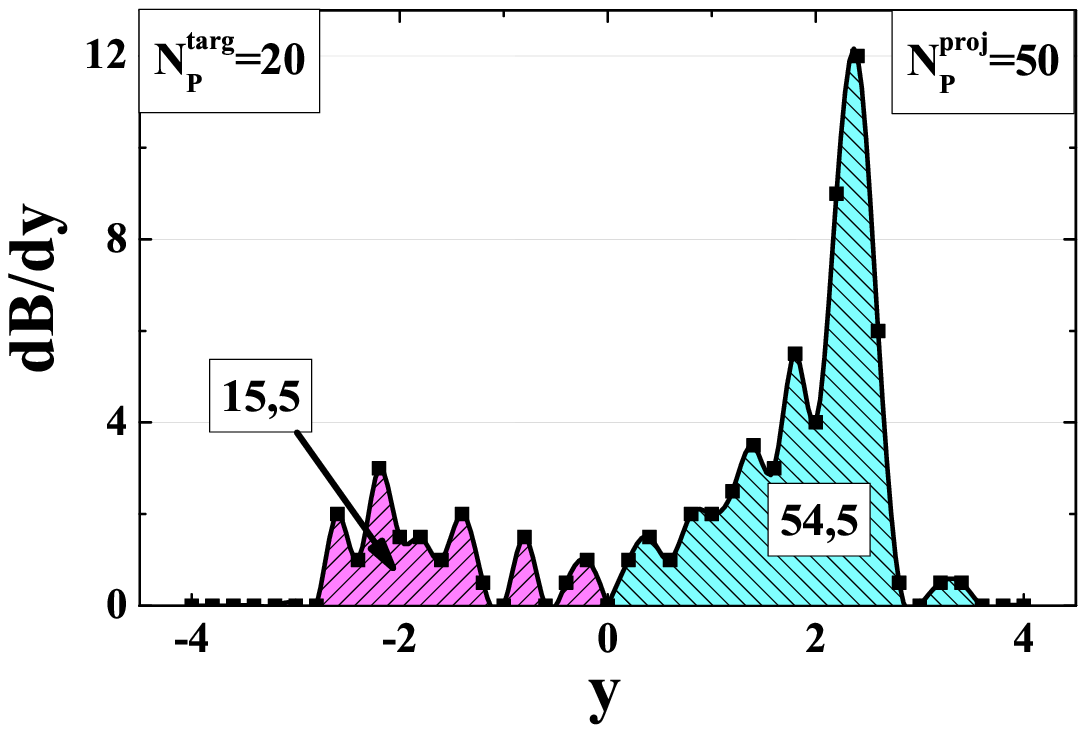,width=8cm} \caption{The HSD results for
Pb+Pb collisions at 158 AGeV for the rapidity distributions of
baryon numbers in nonsymmetric samples with $N_P^{proj}=50,
N_P^{targ}=78$ ({\it left}), and $N_P^{proj}=50, N_P^{targ}=20$
({\it right}). } \label{nocentr}
\end{figure}

The numbers of target and projectile participants are defined as
$N_P^{targ}\equiv A-N_S^{targ}$ and $N_P^{proj}\equiv A-N_S^{proj}$.
The actual event-by-event numbers of baryons in the target and
projectile hemispheres, $N_B^t$ and $N_B^p$, may differ from
$N_P^{targ}$ and $N_P^{proj}$. This is because a transfer of baryons
between the projectile and target hemispheres arises from the
production of baryon-antibaryon pairs. The partners of each newly
created $b\overline{b}$-pair can be detected with non-zero probability
in different hemispheres.
We introduce  $b^t\equiv N_B^t-N_P^{targ}$ and the number of antibaryons
in the target hemisphere, $\overline{b}^t$. Similarly, $b^p\equiv
N_B^p-N_P^{proj}$, while $\overline{b}^p$ is the number of
antibaryons in the projectile hemisphere.  One finds:
\eq{
& \omega_B^t~\equiv
\frac{Var(N_P^{targ}~+~b^t~-~\overline{b}^t)}{\langle B^t\rangle}~
=~\omega_P^{targ}~ \nonumber \\
&+~\frac{1}{N_P^{proj}}~\left[Var(b^t)~+~Var(\overline{b}^t) ~+~
2~\Delta(N_P^{targ},~b^t)~-~2~\Delta(N_P^{targ},~\overline{b}^t)~-~
2~\Delta(b^t,~\overline{b}^t)~\right]~,\label{termsBt}
\\
&\omega_B^p~\equiv
\frac{Var(N_P^{proj}~+~b^p~-~\overline{b}^p)}{\langle B^p\rangle}~
=~\frac{1}{N_P^{proj}}~\left[Var(b^p)~+~Var(\overline{b}^p) ~-~
2~\Delta(b^p,~\overline{b}^p)~\right]~,\label{termsBp}
}
where
\eq{\label{correlator}
\Delta(N_{1},~N_{2})~\equiv~ \langle N_{1}~\cdot~N_2\rangle
~-~\langle N_{1}\rangle ~\cdot~\langle N_{2}\rangle~.
}
As $N_P^{proj}=const$ in the sample, it follows that
$\omega_P^{proj}=0$, $\Delta(N_P^{proj},~b^p)=0$,
$\Delta(N_P^{proj},~\overline{b}^p)=0$, these terms are absent
in the r.h.s. of Eq.~(\ref{termsBp}). Different terms of
Eq.~(\ref{termsBt}) and Eq.~(\ref{termsBp}) found from the HSD
simulations are presented in Fig~\ref{terms}.
One observes that terms of Eq.~(\ref{termsBt},\ref{termsBp})
expressing the fluctuations of antibaryons,
$Var(\overline{b}^p)/N_P^{proj}$, and the correlation terms,
$2\Delta(N_P^{targ},~\overline{b}^t)/N_P^{proj}$ and $~-~
2\Delta(b^t,~\overline{b}^t)/N_P^{proj}$, with antibaryons included,
are small. Therefore, one finds, $\omega_B^p\cong
Var(b^p)/N_P^{proj}$. In the target hemisphere, the
$\omega_P^{targ}$ gives the main contribution to $\omega_B^t$ in
Eq.~(\ref{termsBt}). The term $Var(b^t)/N_P^{proj}$ also contributes
to $\omega_B^t$, similarly to that, $Var(b^p)/N_P^{proj}$, in the
projectile hemisphere. However, the main additional term to
$\omega_B^t$ is $2\Delta(N_P^{targ},~b^t)/N_P^{proj}$, which is due
to (positive) correlations between $N_P^{targ}$ and $b^t$. This
implies that in events with large $N_P^{targ}$ (i.e. $N_P^{targ}
>\langle N_P^{targ}\rangle \cong N_P^{proj}$) some additional
baryons move from the projectile to the target hemisphere, and when
$N_P^{targ}$ is small (i.e. $N_P^{targ} <\langle N_P^{targ}\rangle
\cong N_P^{proj}$) the baryons move in the reverse direction from the
target to the projectile  hemisphere as shown in
Fig.~\ref{nocentr}.

This HSD result looks rather unexpected. We remind that
Eq.~(\ref{alpha1}) predicts for $\omega_B^t$ the opposite behavior:
due to a simple mixing of baryons between the target and projectile
hemispheres the initially large fluctuations, $\omega_P^{targ}$, are
transformed into smaller ones, $\omega_B^t$. It seems that the
origin of this effect is the following: For $N_P^{targ}> N_P^{proj}$
each projectile nucleon interacts, in average, more often than the
target nucleon. The projectile participant loses then a larger part
of its energy, and in the rapidity space its position becomes closer
to $y_{c.m.}=0$ than the position of target participants. This gives
to projectile participants more chances to move due to further
rescatterings from projectile to target hemisphere, in a comparison
with target participants to move in the opposite direction. For
$N_P^{targ}< N_P^{proj}$ there is a reverse situation. This fact was
not taken into account in Eqs.~(\ref{PBt},\ref{PBp}) where it has
been assumed that the mixing probability $\alpha$ is the same for
projectile and target participants, and independent of $N_P^{targ}$.

\section{Net Electric Charge Fluctuations}

\begin{figure}[ht!]
\epsfig{file=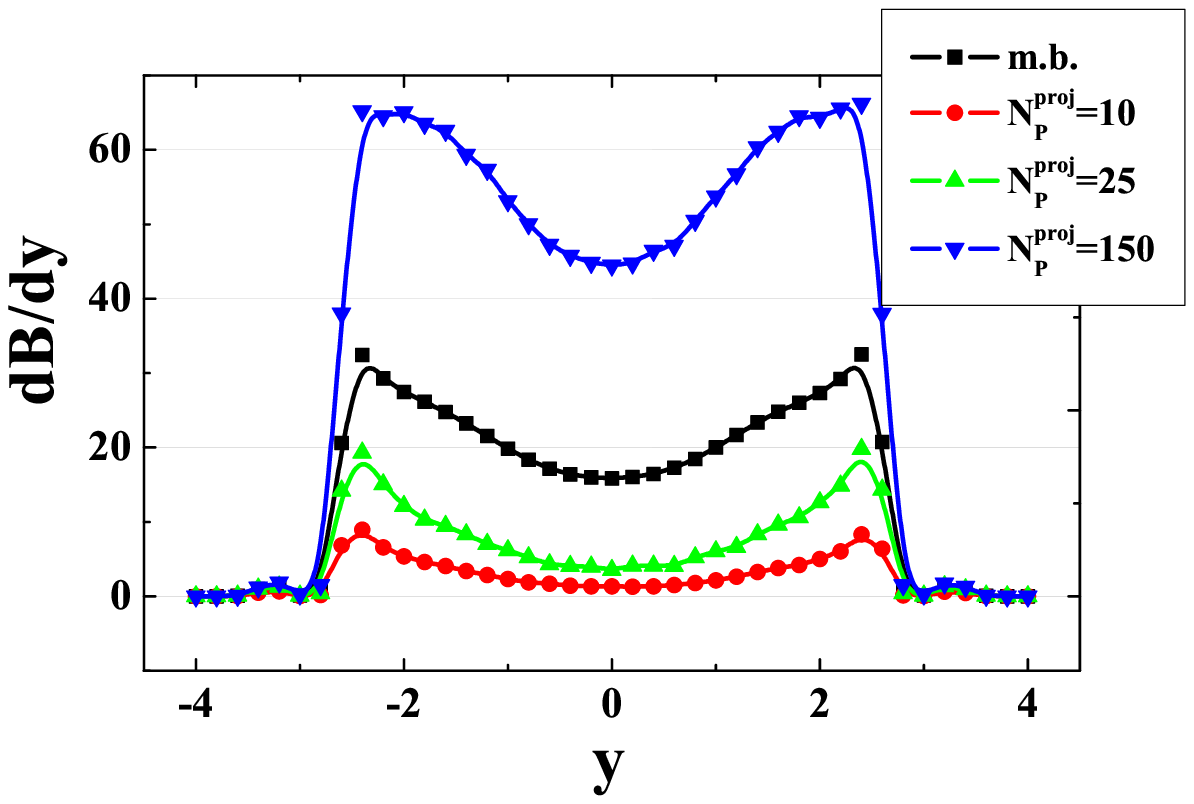,width=8cm}
\epsfig{file=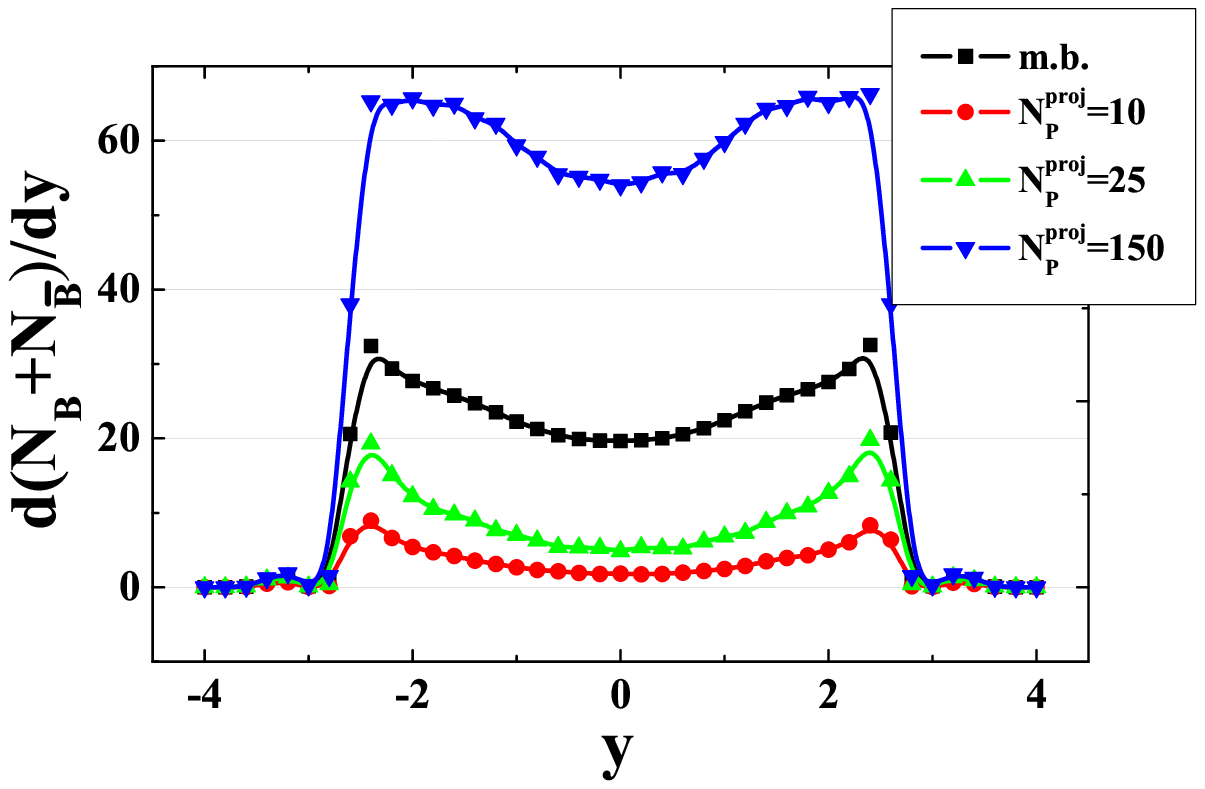,width=8cm} \caption{The HSD rapidity
distributions in Pb+Pb collisions at 158~AGeV for the net baryon
number, $B=N_B-N_{\overline{B}}$ ({\it left}), and total number of
baryons, $N_B+N_{\overline{B}}$ ({\it right}), at different
$N_P^{proj}$ and in the minimum bias (m.b.) sample.} \label{Bt}
\end{figure}

The T-, M- and R-models give very different predictions for
$\omega_B^p$ and $\omega_B^t$ for the samples of events with fixed
values of $N_P^{targ}$. Additional interesting correlations between
the $B^t$ and $B^p$ numbers, as those seen in the HSD simulations,
can be expected. Unfortunately, they may be difficult to test
experimentally as an identification of protons and a measurement of
neutrons in a large acceptance  in a single event is difficult.
Measurements of the charged particle multiplicity in a large acceptance
can be performed with the existing detectors.  In this section we
consider the HSD results for the net electric charge, $Q$,
fluctuations.
As $Q\cong 0.4 B$ in the initial heavy nuclei one can naively expect
that $Q$ fluctuations are quite similar to $B$ fluctuations.  We
stress, however, a principal difference between $Q$ and $B$ in
relativistic A+A collisions. Fig.~\ref{Bt} demonstrates the rapidity
distributions of the net baryon number, $B=N_B-N_{\overline{B}}$~
(left), and {\it total} number of baryons, $N_B+N_{\overline{B}}$~
(right), for different centralities in Pb+Pb collisions at 158~AGeV.
One observes that both quantities are very close to each other; the
$y$-dependence and absolute values are very close for $B$ and
$N_B-N_{\overline{B}}$ distributions. This is, of course, because the
number of antibaryons is rather small, $N_{\overline{B}}\ll N_B$.

\begin{figure}[ht!]
\epsfig{file=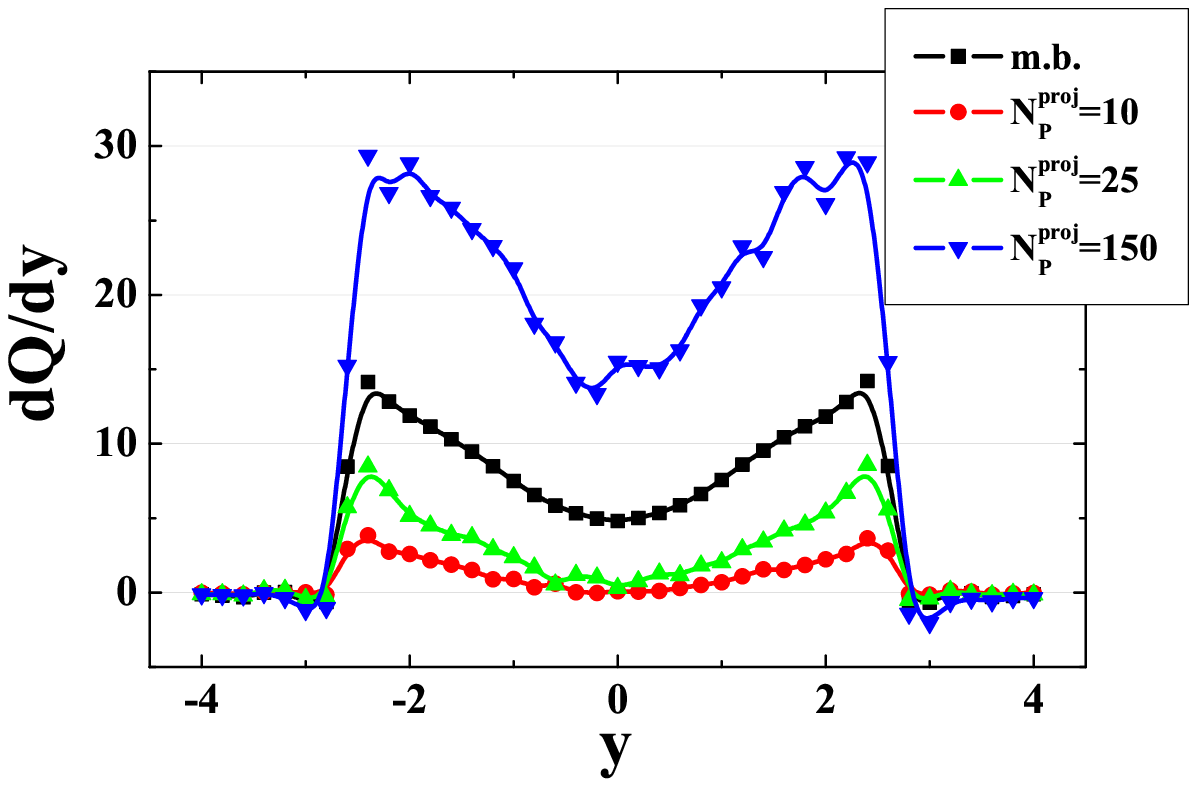,width=8cm}
\epsfig{file=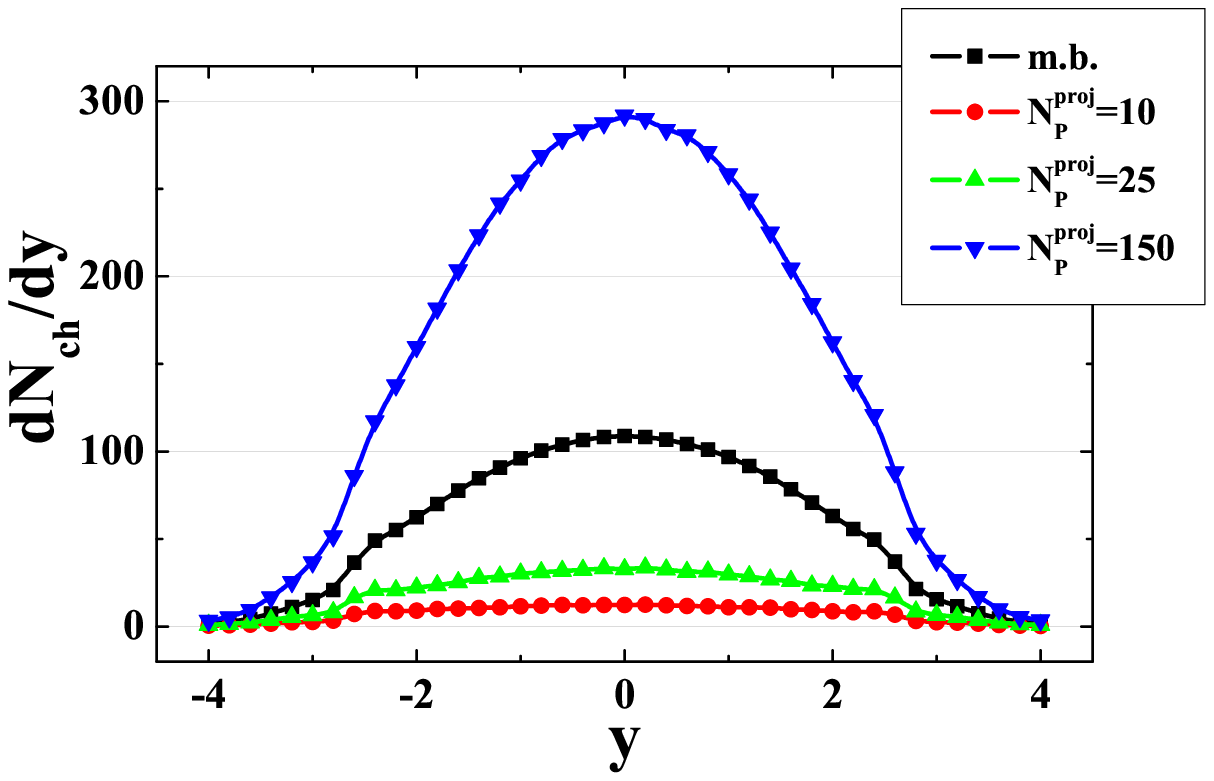,width=8cm} \caption{The same as in
Fig.~\ref{Bt} but for the electric charge  $Q=N_+ -N_{-}$~ ({\it
left}), and total number of charged particles, $N_{ch}\equiv
N_++N_{-}$~ ({\it right}).} \label{Qt}
\end{figure}

Fig.~\ref{Qt} shows the same as Fig.~\ref{Bt} but for the electric
charge  $Q=N_+ -N_{-}$~ (left), and {\it total} number of charged
particles, $N_{ch}\equiv N_++N_{-}$~ (right).
%
The $y$-dependence of $dQ/dy$ and $dN_{ch}/dy$ is quite different.
Besides, the absolute values of $N_{ch}$ are about 10 times larger
than those of $Q$.  This implies that $Q\ll N_+\approx N_-$.

\begin{figure}[ht!]
\epsfig{file=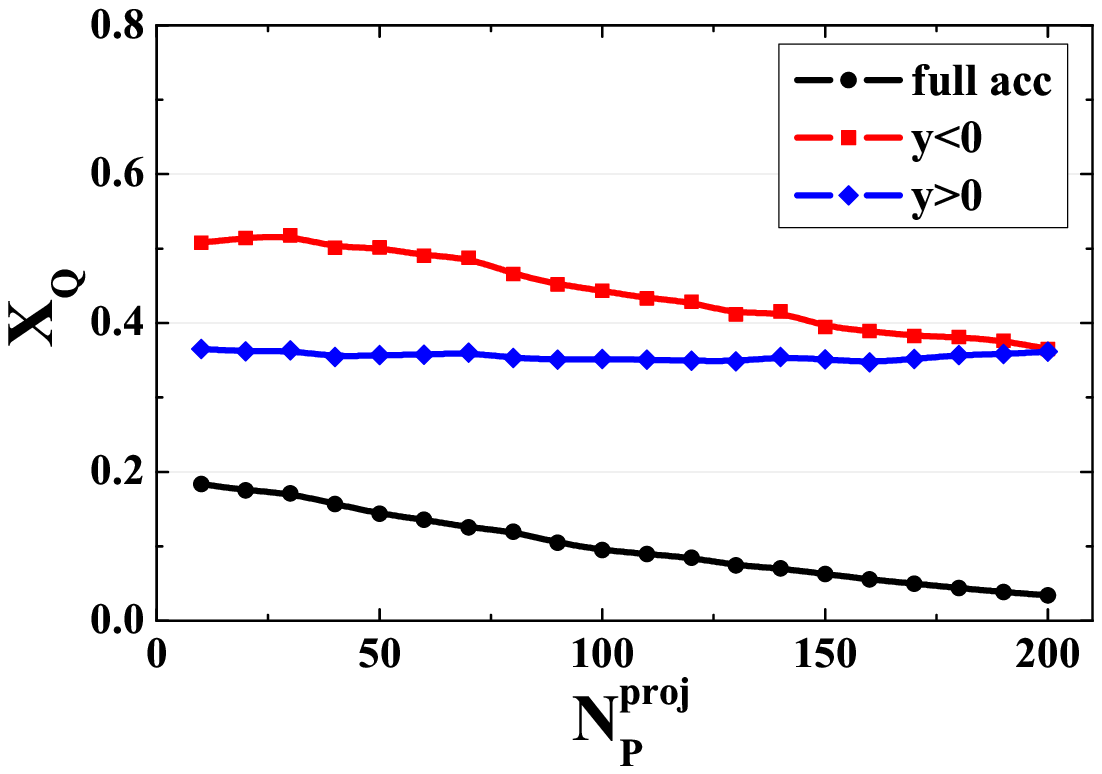,width=8cm}
 \epsfig{file=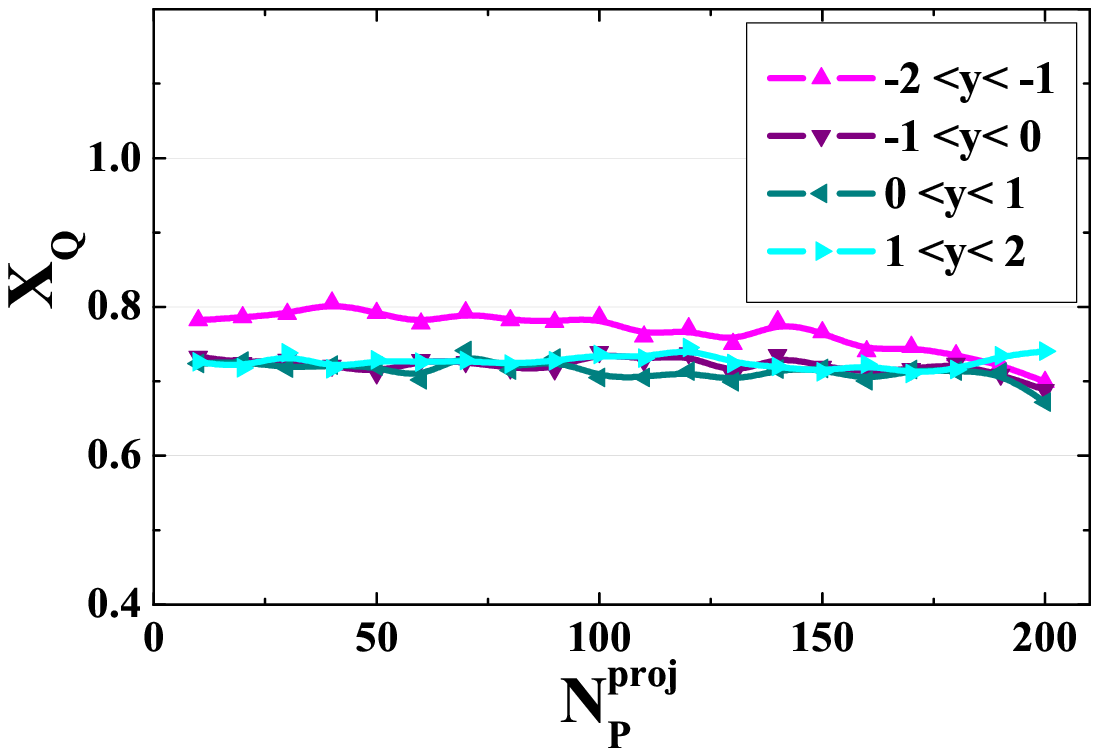,width=8cm}
\caption{{\it Left}: The HSD simulations in Pb+Pb collisions at
158~AGeV for $X_Q$ at different values of $N_P^{proj}$ in the full
acceptance (lower curve), for the projectile (middle curve) and
target (upper curve) hemispheres. {\it Right:} The same, but for
symmetric rapidity intervals in the c.m.s. } \label{xQ1}
\end{figure}

In the previous section we have used the scaled variance $\omega_B$
to quantify the measure of the net baryon fluctuations. It appears
to be a useful variable as $\omega_B$ is straightforwardly connected
to $\omega_P^{targ}$ and due to the relatively small number of
antibaryons. Fig.~\ref{Qt} tells that $\omega_Q$ is a bad  measure
of the electric charge fluctuations in high energy A+A collisions.
One observes that $\omega_Q\equiv Var(Q)/\langle Q\rangle$ is much
larger than 1 simply due to the small value of $\langle Q\rangle$ in
a comparison with $N_+$ and $N_-$. If the A+A collision energy
increases, it follows, $\langle Q\rangle \rightarrow 0$, and thus
$\omega_Q\rightarrow \infty$. The same will happen with $\omega_B$,
too, at much larger  energies. A useful measure of the
net electric charge fluctuations is the quantity (see, e.g.,
\cite{fluc7a}):
\eq{
X_Q~\equiv~\frac{Var(Q)}{\langle N_{ch}\rangle}~. \label{XQ}
}
%
A value of $X_Q$ can be easily calculated for the Boltzmann ideal
gas in the grand canonical ensemble. In this case the number of
negative and positive particles fluctuates according to the Poisson
distribution (i.e. $\omega_-=\omega_+=1$), and the correlation
between $N_+$ and $N_-$ are absent (i.e. $\langle N_+N_-\rangle =
\langle N_+\rangle \langle N_-\rangle$), so that $X_Q=1$. On the
other hand, the canonical ensemble formulation (i.e. when $Q=const$
fixed exactly for all microscopic states of the system) leads to
$X_Q=0$. Fig.~\ref{xQ1} shows the results of the HSD simulations for
the full acceptance, for the projectile and target hemispheres
(left), and also for symmetric rapidity intervals in the c.m.s.
(right).

The $Q$ fluctuation in the full acceptance is due to
$N_P^{targ}$ fluctuations. As $Q\cong 0.4 B$ in colliding (heavy)
nuclei, one may expect $Var(Q)\cong 0.16~ Var(B)$. In addition,
$\langle N_{ch}\rangle \cong 4 \langle N_P \rangle$ at 158~AGeV, so
that one estimates $X_Q \cong 0.04~ \omega_B$ for the fluctuations
in the full phase space. The actual values of $X_Q$ presented in
Fig.~\ref{xQ1} (left) are about 3 times larger. This is because of
$Q$ fluctuations  due to different event-by-event values of proton
and neutron participants even in a sample with fixed values of
$N_P^{proj}$ and $N_P^{targ}$.

\begin{figure}[ht!]
\epsfig{file=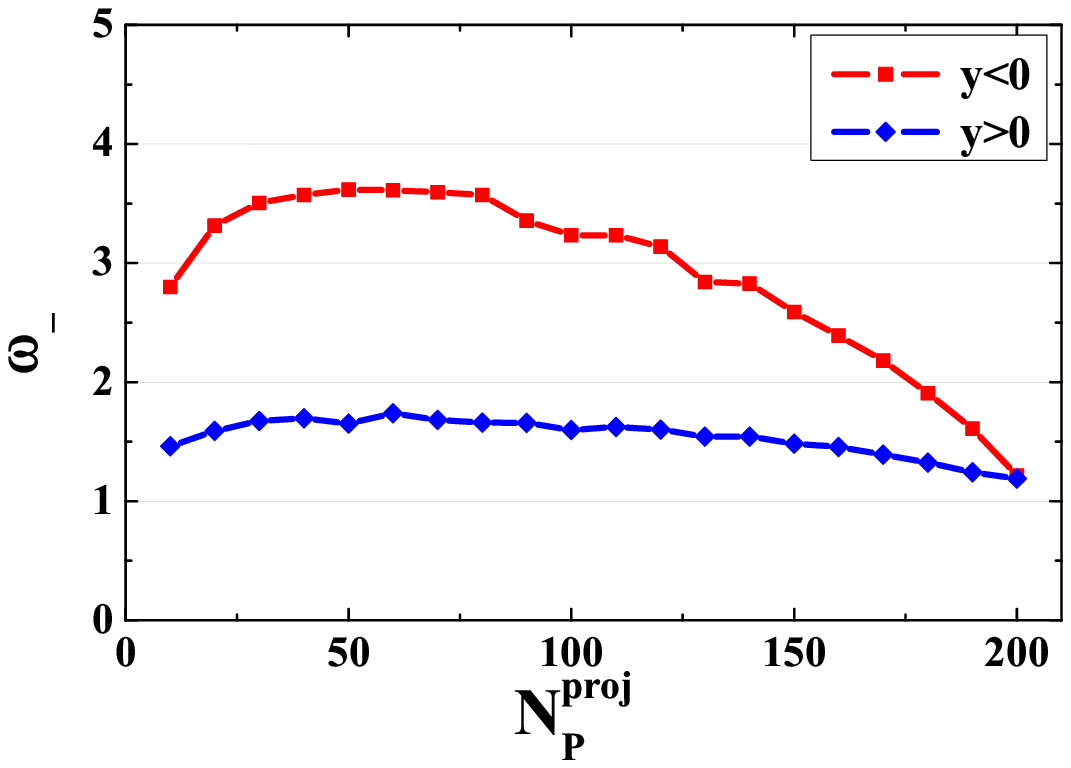,width=8cm}
 \epsfig{file=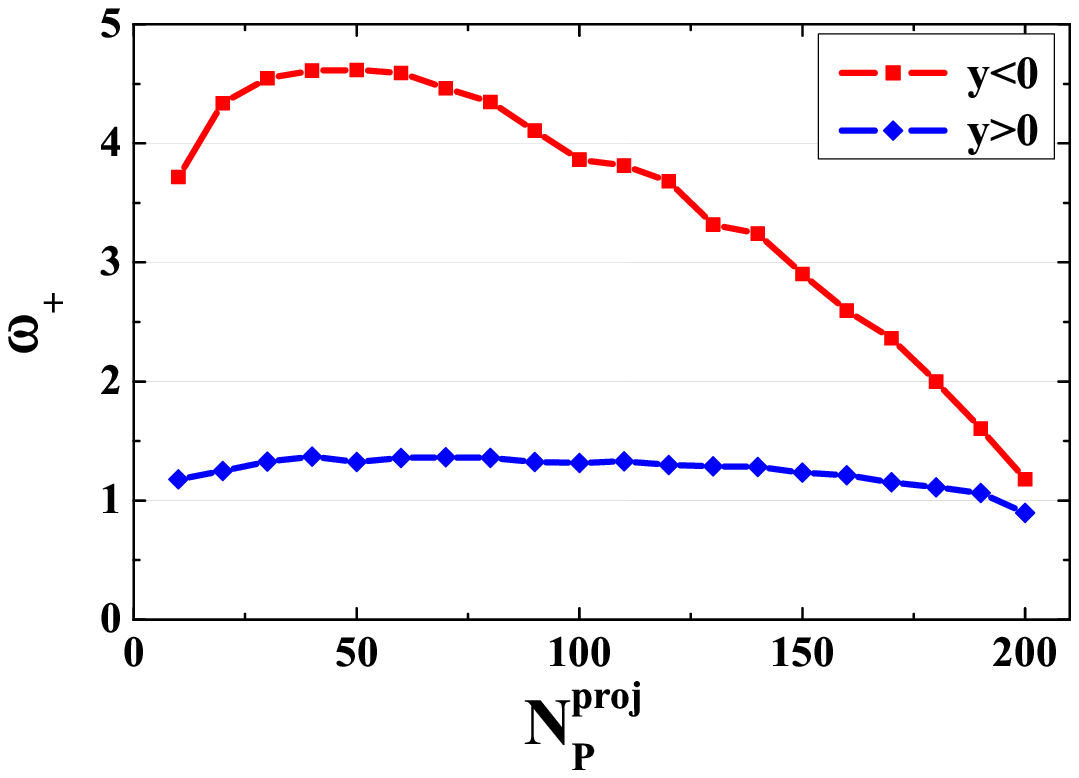,width=8cm}
\caption{The HSD results for the scaled variances of negatively
({\it left}) and positively ({\it right}) charged hadrons in Pb+Pb
collisions at 158~AGeV for the projectile (lower curves) and target
(upper curves) hemispheres. } \label{omegapm}
\end{figure}

From Fig.~\ref{xQ1} (right) one sees only a tiny difference between
the $X_Q$ values in the symmetric rapidity intervals in the
projectile and target hemispheres, and slightly stronger effects for
the whole projectile and target hemispheres (Fig.~\ref{xQ1}, right).
In fact, the fluctuations of $N_+$ and $N_-$ are very different in
the projectile and target hemispheres, and the scaled variances
$\omega_{+}^t$ and $\omega_-^t$ have a very strong
$N_P^{proj}$-dependence. This is shown in Fig.~\ref{omegapm}
obtained in our previous study \cite{KGB}.

\begin{figure}[ht!]
\epsfig{file=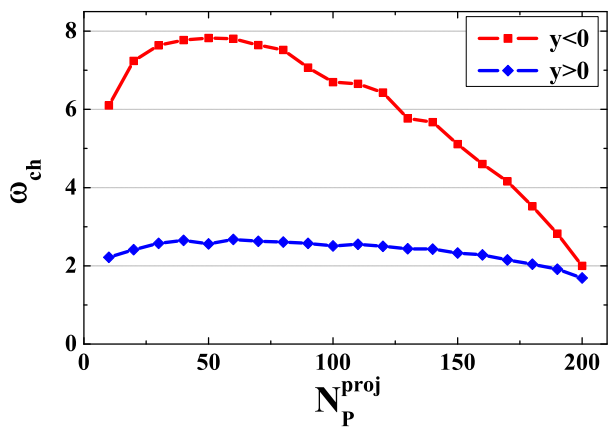,width=8cm} \caption{The HSD results for the
scaled variances of all charged hadrons, $\omega_{ch}$, in Pb+Pb
collisions at 158~AGeV for the projectile (lower curve) and target
(upper curve) hemispheres.} \label{omegach}
\end{figure}

 The $X_Q$ can be presented in two equivalent forms
\eq{
X_Q~=~ \omega_+~\frac{\langle N_+\rangle }{\langle N_{ch}\rangle
}~+~\omega_-~\frac{\langle N_-\rangle}{\langle N_{ch}\rangle}~-~
2~\frac{\Delta(N_+,N_-)}{\langle N_{ch}\rangle
}~=~2~\omega_+~\frac{\langle N_+\rangle}{\langle N_{ch}\rangle}~+~
2~\omega_-~\frac{\langle N_-\rangle}{\langle
N_{ch}\rangle}~-~\omega_{ch}~.\label{XQ-1}
}
Eq.~(\ref{XQ-1}) is valid for any region of the phase space: full
phase space, projectile or target hemisphere, etc. As seen from
Fig.~\ref{omegapm}, both $\omega_+^t$ and $\omega_-^t$ are large and
strongly $N_P^{proj}$-dependent. This is not seen in $X_Q^t$ because
of strong correlations between $N_+^t$ and $N_-^t$, i.e. the term
$2~\Delta(N_+,N_-)/\langle N_{ch}\rangle $ compensates $\omega_+$
and $\omega_-$ terms in Eq.~(\ref{XQ-1}). This is also seen from
Fig.~\ref{omegach}.
%
%
A cancellation of strong $N_P^{proj}$-dependence in the target
hemisphere takes place between the sum of $\omega_+^t$ and
$\omega_-^t$ terms of Eq.~(\ref{XQ-1}), and the
$~\omega_{ch}^t$-term.

\begin{figure}[ht!]
\epsfig{file=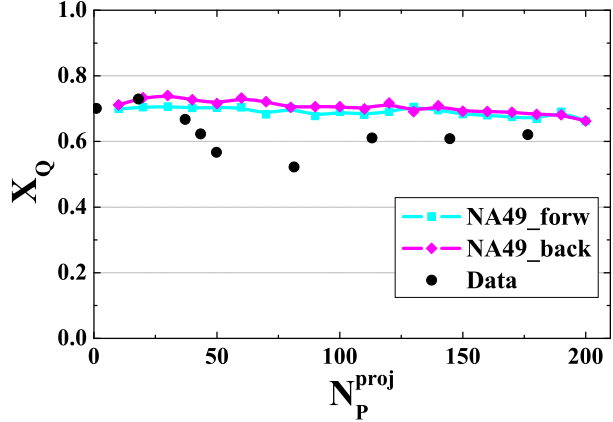,width=8cm} \caption{The HSD results for $X_Q$
for Pb+Pb collisions at 158~AGeV for the forward rapidity interval
$1.1<y<2.6$ inside the projectile hemisphere. The solid dots are the
estimates obtained from Eq. (\protect\ref{XQ-1}) using the NA49
experimental data \protect\cite{NA49} (the errorbars are not
indicated here). For  illustration, the HSD results in the symmetric
backward rapidity interval $-2.6<y<-1.1$ (target hemisphere) are
also presented.} \label{xQdata}
\end{figure}

Fig.~\ref{xQdata} shows a comparison of the HSD results for $X_Q$
with NA49 data in Pb+Pb collisions at 158~AGeV for the forward
rapidity interval $1.1<y<2.6$ inside the projectile hemisphere with
additional $p_T$-filter imposed.
As an illustration, the HSD results in the symmetric backward rapidity
interval $-2.6<y<-1.1$ (target hemisphere) are also included. One
observes no difference between the $X_Q$ results for the NA49
acceptance in the projectile and target hemispheres. The HSD values for
$\omega_+$, $\omega_-$, and $\omega_{ch}$ are rather different in the
projectile and target hemispheres for the NA49 acceptance (see
Figs.~\ref{omegapm} and \ref{omegach}). This is not seen in
Fig.~\ref{xQdata} for $X_Q$. As explained above a cancellation between
$\omega_+$, $\omega_-$ and $\omega_{ch}$ terms take place in
Eq.~(\ref{XQ-1}). In fact, NA49 did not perform the $X_Q$ measurements.
The $X_Q$-data (solid dots) presented in Fig.~\ref{xQdata} are obtained
from Eq.~(\ref{XQ-1}) using the NA49 data for $\omega_+$, $\omega_-$,
and $\omega_{ch}$ as well as $\langle N_+\rangle$, $\langle
N_-\rangle$, and $\langle N_{ch}\rangle$ \cite{NA49}. Such a procedure
leads, however, to very large errors  for $X_Q$ (which are not
indicated in Fig. \ref{xQdata}) which excludes any conclusion
about the (dis)agreement of HSD results with NA49 data.

\section{Fluctuations in Most Central Collisions}

\begin{figure}[ht!]
\epsfig{file=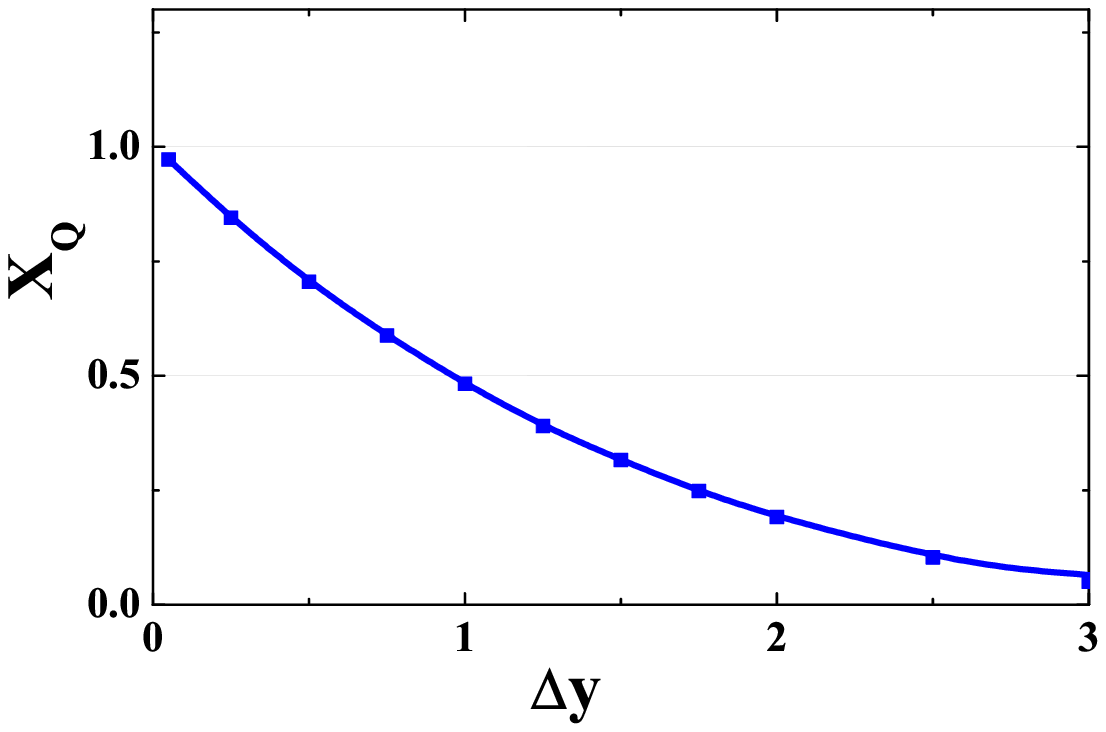,width=8cm}
\epsfig{file=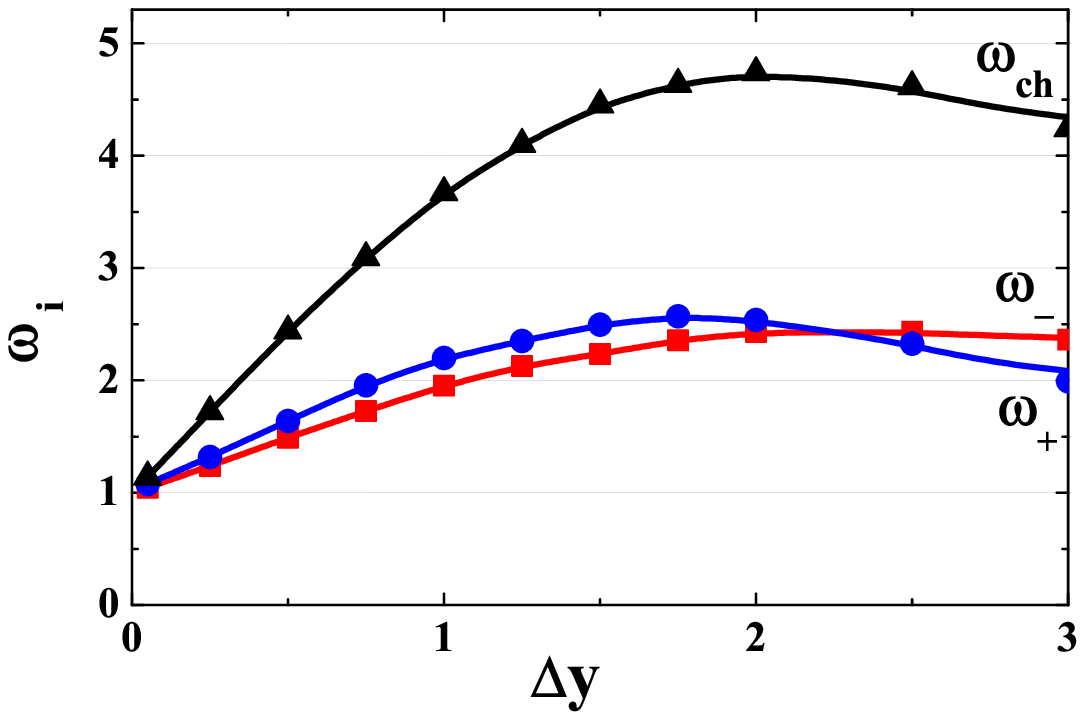,width=8cm}\caption{The HSD results for
electric charge fluctuations in 2\% most central Pb+Pb collisions at
158~AGeV in the symmetric rapidity interval                       ,
$\Delta Y = [-y, y]$ as a function of  $\Delta y= \Delta Y/2$
in the c.m.s. A {\it left} panel shows the behavior of $X_Q$, and a
{\it right} one demonstrates separately $\omega_+$, $\omega_-$, and
$\omega_{ch}$. } \label{xQ-dY}
\end{figure}

In this section we consider the baryon number and electric charge
fluctuations in the symmetric rapidity interval $[-y, y]$ in
the c.m.s. for the most central Pb+Pb events. We chose the
sample of most central events by restricting the impact parameter
to $b<2$~fm. It gives about 2\%  most central Pb+Pb
collisions from the whole minimum bias sample. Fig.~\ref{xQ-dY}
shows the HSD results for electric charge fluctuations in 2\%
most central Pb+Pb collisions  for the symmetric rapidity interval
$\Delta Y = [-y, y]$ in the c.m.s. as the function of $\Delta y =\Delta Y/2$.
For $\Delta Y\rightarrow 0$ one finds $X_Q\rightarrow 1$. This can
be understood as follows: For $\Delta Y\rightarrow 0$ the
fluctuations of negatively, positively and all charged particles
behave as for the Poisson distribution: $\omega_+\cong \omega_-\cong
\omega_{ch}\cong 1$. Then from Eq.~(\ref{XQ-1}) it follows that
$X_Q\cong 1$, too. From Fig.~\ref{xQ-dY} (right) one observes that
$\omega_+$, $\omega_-$, and $\omega_{ch}$ all increase with
increasing interval $\Delta Y$. However, $X_Q$
decreases with $\Delta Y$ and -- because of global $Q$ conservation --
it goes approximately to zero when all final particles are accepted.

\begin{figure}[ht!]
\epsfig{file=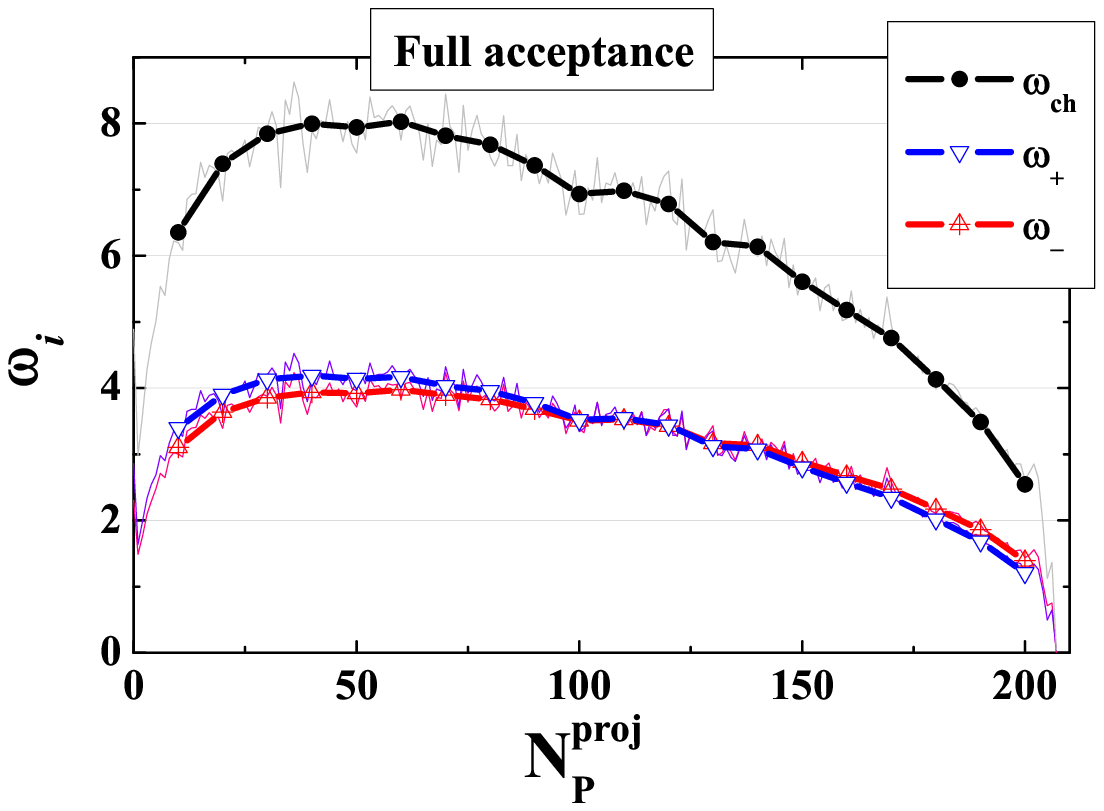,width=8cm}
\epsfig{file=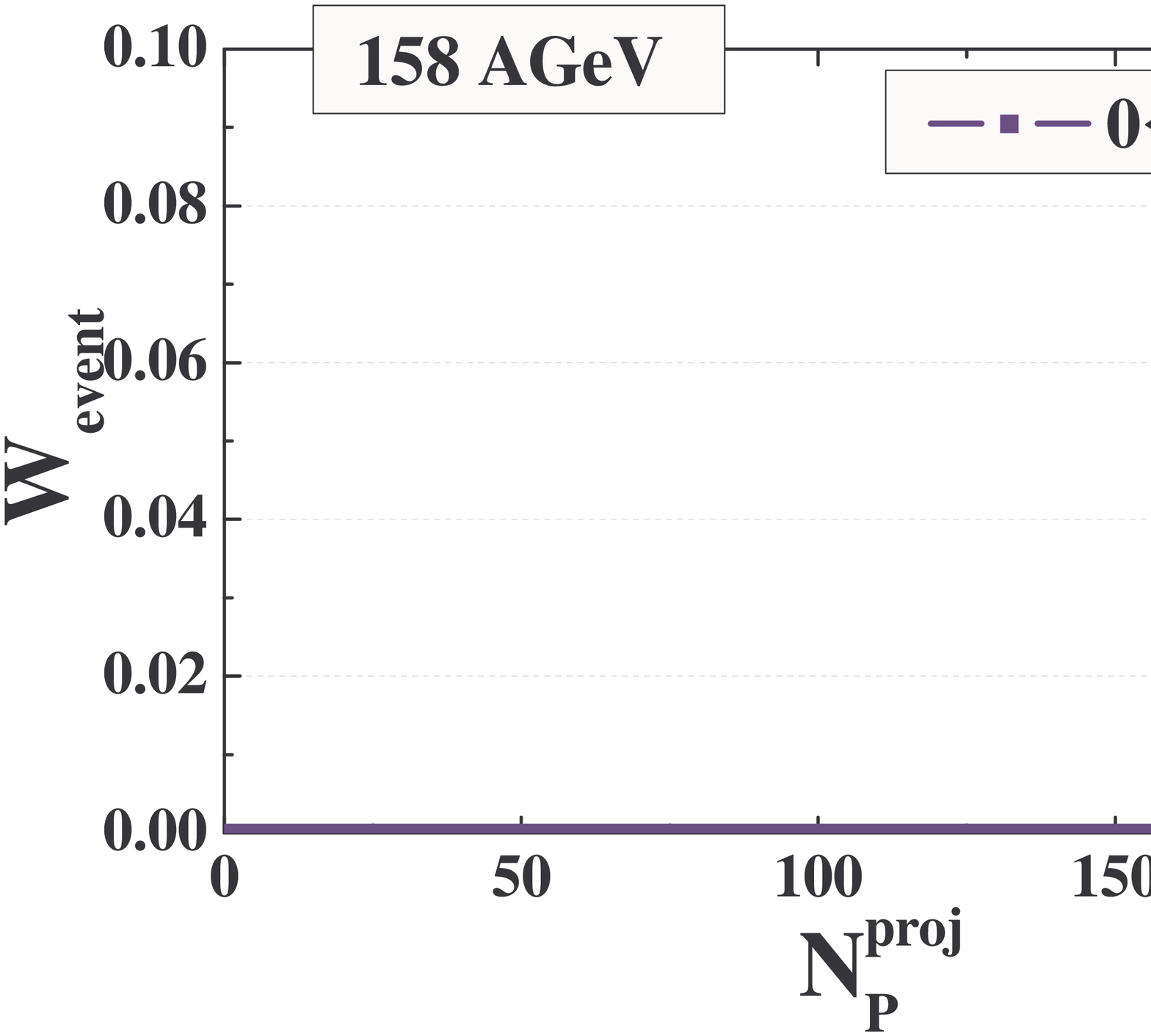,scale=0.3} \hfill\phantom{a}
\caption{The HSD results in Pb+Pb
collisions at 158 AGeV. {\it Left:} The scaled variances $\omega_+$,
$\omega_-$, and $\omega_{ch}$ in the full acceptance. {\it Right:}
The distributions of events over $N_P^{proj}$ in most central
collisions with $b<2$~fm. } \label{omegai-full}
\end{figure}

In Fig.~\ref{omegai-full} (left) the HSD results for the scaled
variances are presented in full acceptance as functions of
$N_P^{proj}$. Fig.\ref{omegai-full} (right) demonstrates the
probability distribution of events with $b<2$~fm over $N_P^{proj}$.
%
One observes that even in the 2\% centrality sample the values of
$N_P^{proj}$ are noticeably smaller than the maximum value, $A=208$.
As seen from Fig.~\ref{omegai-full} (left) the HSD values of
$\omega_+$, $\omega_-$, and $\omega_{ch}$ become then essentially
larger than 1 in agreement with those presented in
Fig.~\ref{xQ-dY}.

\begin{figure}[ht!]
\epsfig{file=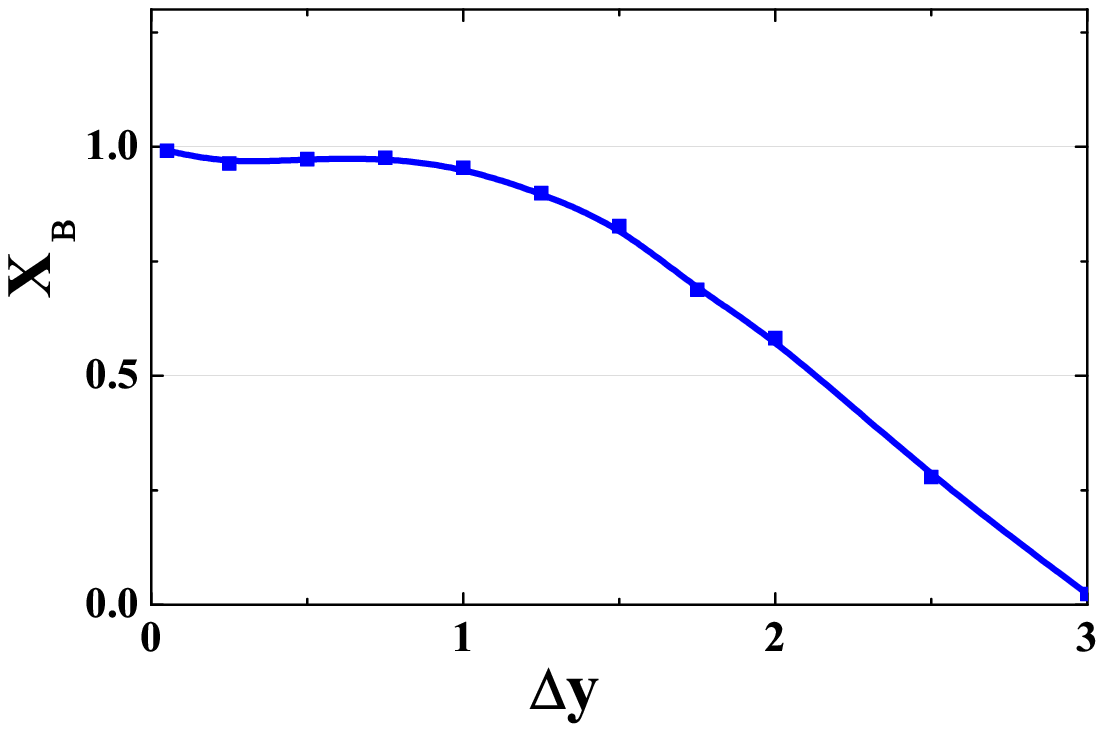,width=8cm}
\epsfig{file=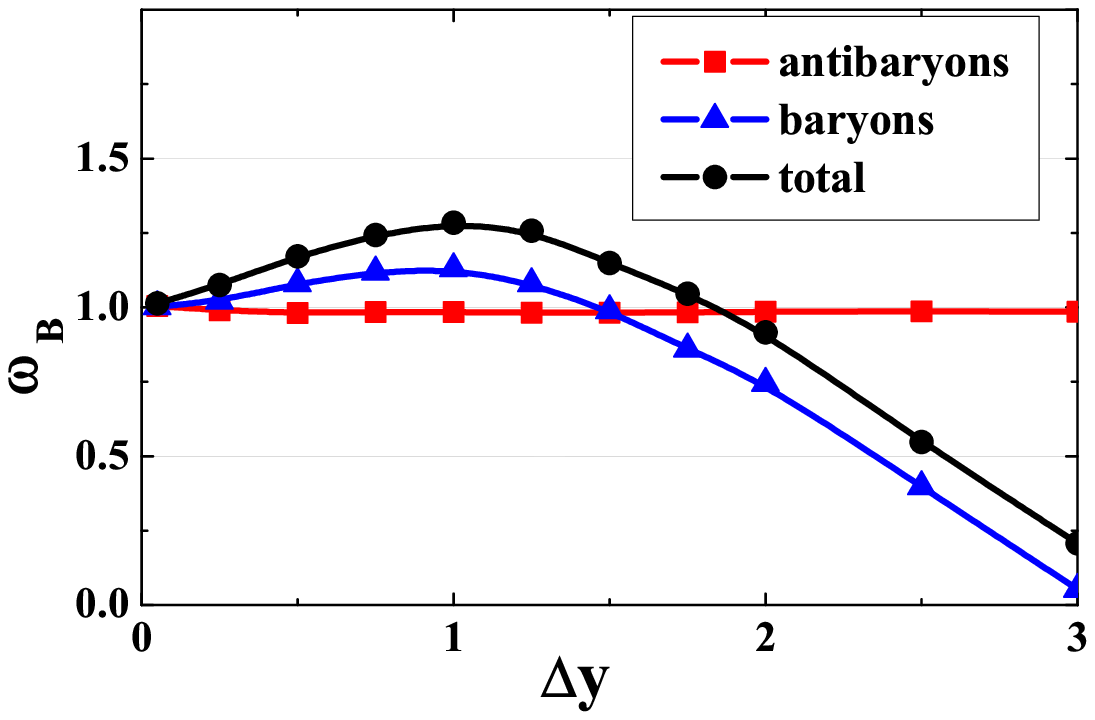,width=8cm} \caption{The HSD results for net
baryon number fluctuations in 2\%  most central Pb+Pb collisions at
158~AGeV in the symmetric rapidity interval $\Delta Y = [-y, y]$
as a function of  $\Delta y= \Delta Y/2$
in the c.m.s. The {\it left} panel shows the behavior of $X_B$, and
a {\it right} panel presents separately $\omega_{N_B}$,
$\omega_{N_{\overline{B}}}$, and $\omega_{N_B+N_{\overline{B}}}$. }
\label{xB-dY}
\end{figure}

 Fig.~\ref{xB-dY} shows the net baryon number
fluctuations in the symmetric rapidity interval $[-y, y]$
in the c.m.s. as the function of $\Delta Y$.
As a measure of the net baryon number fluctuations we have used
the quantity,
\eq{\label{XB}
X_B~\equiv~\frac{Var(B)}{\langle N_B~+~N_{\overline{B}}\rangle }~.
}
As for the electric charge, one finds that $X_B \rightarrow 1$ at
$\Delta Y\rightarrow 0$ (this is because all $\omega_{N_B}$,
$\omega_{N_{\overline{B}}}$, and $\omega_{N_B+N_{\overline{B}}}$ go
to 1 in this limit  (see Fig.~\ref{xB-dY}, left), and $X_B
\rightarrow 0$ at upper limit of $\Delta Y$ because of global baryon
number conservation.

Writing the variance $Var(B)$ in the form,
\eq{
 Var(B)~=
2~Var(N_B)~+~2~Var(N_{\overline{B}})~-~Var(N_B~+~N_{\overline{B}})~,
\label {VarB}
}
 we find
\eq{\label{XB1}
 X_B~=~ 2~\omega_{N_B}~\frac{\langle N_B\rangle}{\langle
N_B~+~N_{\overline{B}\rangle}}~+~
2~\omega_{N_{\overline{B}}}~\frac{\langle
N_{\overline{B}}\rangle}{\langle
N_B~+~N_{\overline{B}}\rangle}~-~~\omega_{N_B+N_{\overline{B}}}.
}
The behavior of the different terms in Eq.~(\ref{XB1}) is the following:
As seen from Fig.~\ref{xB-dY}, right,
$\omega_{N_{\overline{B}}}\cong 1$ for all values of $\Delta Y$.
This is because $N_{\overline{B}}\ll N_B$, and baryon number
conservation does not affect the fluctuations of antibaryons. Due to
the small number of antibaryons in comparison to baryons,  one also
observes $\omega_B\cong \omega_{N_B}\cong
\omega_{N_B+N_{\overline{B}}}$.

\section{Electric charge fluctuations in central Pb+Pb collisions
at 20, 30, 40, 80 and 160 A GeV}

In this section we present the HSD results for the event-by-event
electric charge fluctuations as measured by the NA49 Collaboration in
central Pb+Pb
collisions at 20, 30, 40, 80 and 160 A GeV \cite{exFq_NA49}. The
interest in this observable (as a signal of deconfinement) is related to
the predicted in Refs.\cite{Ko.1,As.1} suppression of event-by-event
fluctuations of the electric charge in a quark-gluon plasma relative to a
hadron gas. However, these predictions were based on the assumption
that the initial electric charge fluctuations survive the
hadronization phase.

The first experimental measurement of charge fluctuations in
central heavy-ion collisions by PHENIX \cite{Adcox:2002mm}
and STAR \cite{Adams:2003st} at RHIC and by the NA49
\cite{exFq_NA49} at SPS showed a quite moderate suppression of
the electric charge fluctuations. This observation has been attributed to
the fact that the initial fluctuations are distorted by the
hadronization.  In particular, the observed fluctuations might be
related to the final resonance decays.

In this respect it is important to compare the experimental data
with the results of microscopic transport models such as HSD where the
resonance decays are included by default.
In order to quantify the event-by-event electric charge fluctuations
we have calculated the quantity $\Phi$ defined as \cite{exFq_NA49,Ma.1}:
\begin{eqnarray}\label{Phi}
\Phi_q = \sqrt{ \frac{\langle Z^2 \rangle}{\langle N \rangle }} -
\sqrt{\overline{z^2}} \; ,
\end{eqnarray}
where
\begin{equation}\label{z}
z = q - \overline{q},\;\;\;\;\;\;\;\;
Z = \sum_{i=1}^{N}(q_i - \overline{q}) .
\end{equation}
Here $q$ denotes a single particle variable, i.e. electric charge $q$;
$N$ is the number of  particles of the event
within the acceptance, and over-line and $\langle ... \rangle$
denote averaging over
a single particle inclusive distribution and over events, respectively.
By construction,  $\Phi$ of the system, which is an
independent sum of identical sources of particles,  is equal to the
$\Phi$ for a single  source  \cite{Ma.1,Ma.2}.

In order to remove the sensitivity of the final signal to the
trivial global charge conservation (GCC) the measure $\Delta\Phi_{q}$ is
defined as the difference:
\begin{equation}\label{deltaPhi}
\Delta \Phi_q =\Phi_q-\Phi_{q,\rm{GCC}}\;.
\end{equation}
Here the value of $\Phi_q$ is given by \cite{Za.1,Mrowczynski:2001mm}:
\begin{equation}\label{Phiq1}
\Phi_{q,\rm{GCC}}=\sqrt{1-P}-1 ,
\end{equation}
where
\begin{equation}\label{P}
P=\frac {\langle N_{ch}\rangle }{\langle N_{ch} \rangle _{tot}}
\end{equation}
with $\langle N_{ch}\rangle$ and $\langle N_{ch} \rangle _{tot}$
being the mean charged multiplicity in the detector
acceptance and in full phase space (excluding spectator nucleons),
respectively.

By construction, the value of $\Delta \Phi_q$ is zero
if the particles are correlated by global charge conservation only.
It is negative in case of an additional correlation between positively
and negatively charged particles, and it is positive if the positive
and negative particles are anti-correlated \cite{Za.1}.

\begin{figure}[ht!]
\centerline{\epsfig{file=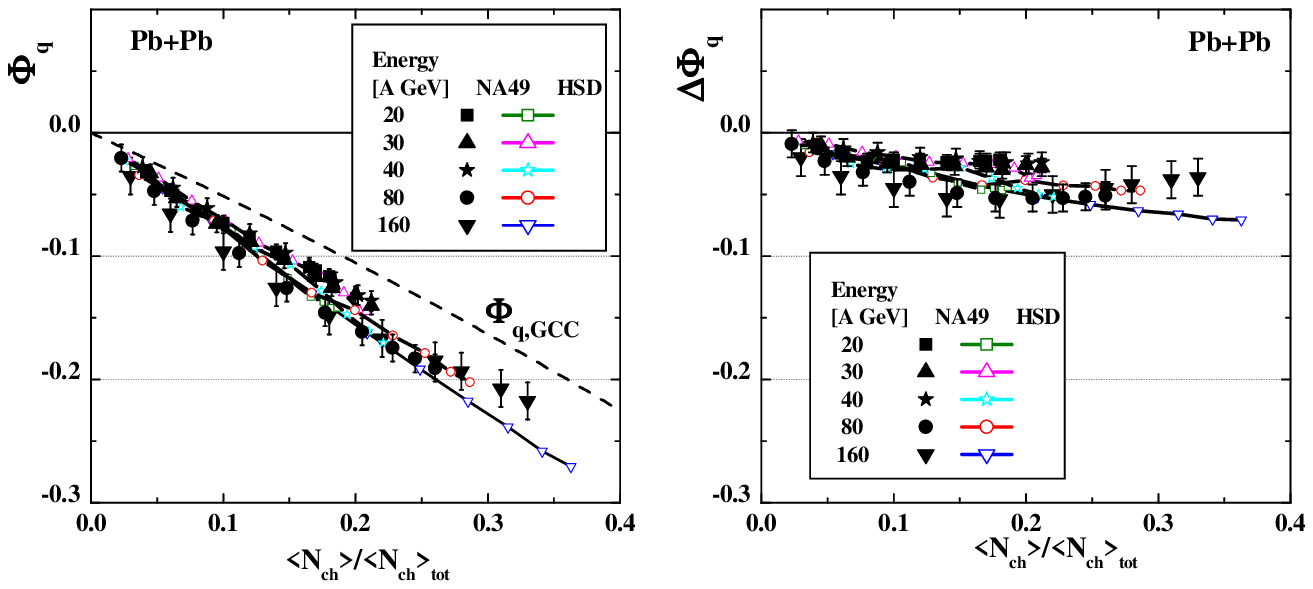,width=15cm}}
\caption{The dependence of the $\Phi_q$ (l.h.s.) and
$\Delta\Phi_q$ (r.h.s.) on the fraction
of accepted particles for central Pb+Pb collisions at 20-158~{\it A}GeV.
The NA49 data \cite{exFq_NA49} are shown as full symbols, whereas the
open symbols (connected by lines) stay for the HSD results.  The
dashed line shows the dependence expected for the case if the only
source of particle correlations is the global charge conservation
$\Phi_{q,GCC}$, Eq.~\ref{Phiq1}.  }
\label{nch}
\end{figure}

Figure \ref{nch} shows the HSD results for the dependence of
$\Phi_q$ (l.h.s.) and $\Delta\Phi_q$ (r.h.s.) on the fraction
of accepted particles
$\langle N_{ch}\rangle$ and $\langle N_{ch} \rangle _{tot}$
(calculated for ten different rapidity intervals increasing in size from
$\Delta y = 0.3$ to $\Delta y = 3$ in equal steps)
for central Pb+Pb collisions at 20, 30, 40, 80 and 158~A GeV.
The NA49 data \cite{exFq_NA49} are shown as full symbols, whereas
the open symbols (connected by lines) reflect the HSD results.  The
dashed line shows the dependence expected for the case if the only
source of particle correlations is the global charge conservation
$\Phi_{q,GCC}$ (Eq. (\ref{Phiq1})).

The data as well as the HSD results for $\Phi_q$ (Fig. \ref{nch}, l.h.s.)
are in a good agreement and
show a monotonic decrease with increasing fraction of accepted particles.
After substraction the contribution by global charge conservation
(the dashed line in Fig.~\ref{nch}), the values
of $\Delta\Phi_q$ vary between $0$ and  $-0.05$ which
are significantly larger than the values expected for QGP fluctuations
($-0.5<\Delta\Phi_q < -0.15$ \cite{Za.1}).

\vspace*{1cm}
\begin{figure}[ht!]
\centerline{\epsfig{file=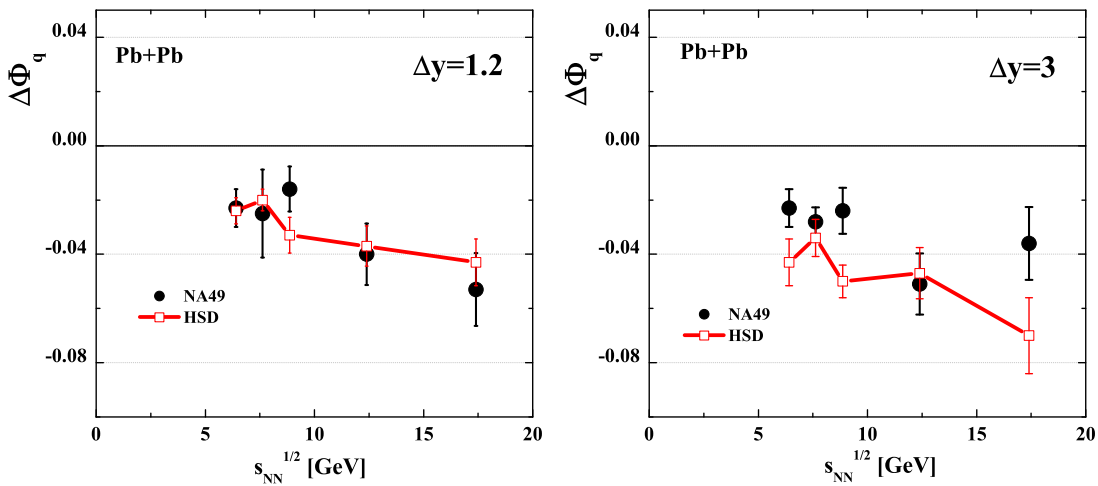,width=15cm}}
    \caption{The energy dependence of $\Delta\Phi_{q}$
measured in central Pb+Pb collisions  for
a narrow rapidity interval $\Delta y = 1.2$
(l.h.s.) and a broad rapidity interval $\Delta y = 3$ (r.h.s.).
The NA49 data \cite{exFq_NA49} are shown as full symbols, whereas the
the open symbols (connected by lines) reflect the HSD results.}
    \label{AllBig}
\end{figure}

Figure \ref{AllBig} presents the energy dependences of $\Delta\Phi_q$
for two selected rapidity intervals -- the intermediate rapidity
interval $\Delta y = 1.2$ (l.h.s.) and for the largest rapidity
interval $\Delta y = 3$ (r.h.s.).
The both, data and HSD results, show the a weak decrease of
$\Delta\Phi_q$ with increasing energy.

The fact that the HSD model, that includes no explicit phase
transition, describes the experimental data can be considered an
independent proof that the event-by-event charge fluctuations are
driven by the hadronization phase and dominantly by the resonance decays
(which are naturally included in HSD) and no longer sensitive to the
initial phase fluctuations from a QGP.

\section{Summary and conclusions}

The goal of this study was to investigate the sensitivity of
event-by-event fluctuations of baryon number and electric charge
to the early stage dynamics of hot and dence nuclear matter
created in heavy-ion collisions at SPS energies
and the influance of the futher hadronization and rescattering phase.
For that perpose we have explored the microscopic HSD transport model
which allows also to investigate (on event-by-event basis) the influence
of the experimental acceptance and the set-up on the final observables.

It has been found that the fluctuations in the number of target
participants strongly influences the baryon number and charged
multiplicity fluctuations.  The consequences of this fact depend
crucially on the dynamics of the initial flows of the conserved
charges and inelastic energy.

For a better quantitative understanding of the microscopic transport
model (HSD) results we have considered 3 limiting groups of models for
nucleus-nucleus collisions:  transparency, mixing and reflection. These
"pedagogical" considerations indicate that the HSD model (as well as
UrQMD, cf. Ref. \cite{KGB}) shows only a small mixing on initial baryon
flow and is  closer to the T-model.  This supports the findings from
Ref. \cite{Weber} about the influence of the partonic degrees of
freedom on the initial phase dynamics which might increase the mixing
by additional strong parton-parton interactions.  Thus, the measurement
of the net baryon number fluctuations helps to quantify the mixing of
initial baryon flow.

The first microscopic event-by-event calculations of the charge
fluctuations $\Delta\Phi_q$ within the HSD model show a good agreement
with the NA49 data at SPS energies.  Thus, this observable is dominated
by the final stage danymics, i.e. the hadronization phase and the
resonance decays, and rather insensitive to the  initial QGP dynamics.

\begin{acknowledgments}
We like to thank W. Cassing, M. Ga\'zdzicki, B.~Lungwitz,
I.N.~Mishustin, St.~Mr\'owczy\'nski, M.~Rybczy\'nski, and
L.M.~Satarov for numerous discussions. The work was supported in
part by US Civilian Research and Development Foundation (CRDF)
Cooperative Grants Program, Project Agreement UKP1-2613-KV-04.
\end{acknowledgments}


%

\end{document}